\documentclass[prodmode,acmtoplas]{acmsmall}

\usepackage[ruled]{algorithm2e}
\usepackage{amsmath}
\usepackage{amssymb}

\renewcommand{\algorithmcfname}{ALGORITHM}
\SetAlFnt{\small}
\SetAlCapFnt{\small}
\SetAlCapNameFnt{\small}
\SetAlCapHSkip{0pt}
\IncMargin{-\parindent}

\acmVolume{34}
\acmNumber{4}
\acmArticle{17}
\acmYear{2012}
\acmMonth{12}

\def \Rm#1{\mbox{\rm #1}}
\def \lsem      {\raise1pt\hbox{\Rm {[\kern-.12em[}}}
\def \rsem      {\raise1pt\hbox{\Rm {]\kern-.12em]}}}
\def \sem#1{\mbox{\lsem$#1$\rsem}}

\newcommand {\qc}[1] {{\sf{#1}}}
\def\>{\ensuremath{\rangle}}
\def\<{\ensuremath{\langle}}
\def\h{\ensuremath{\mathcal{H}}}

\def\g{\ensuremath{\mathcal{G}}}
\def\lh{\ensuremath{\mathcal{L(H)}}}
\def\dh{\ensuremath{\mathcal{D(H})}}

\def\r{\ensuremath{\mathcal{R}}}
\def\m{\ensuremath{\mathcal{M}}}
\def\u{\ensuremath{\mathcal{U}}}

\def\ra{\ensuremath{\rightarrow}}

\def\k{\ensuremath{\mathcal{K}}}
\def\e{\ensuremath{\mathcal{E}}}
\def\f{\ensuremath{\mathcal{F}}}
\def\c{\ensuremath{\mathcal{C}}}
\def\d{\ensuremath{\mathcal{D}}}
\def\dh{\ensuremath{\mathcal{D(H)}}}
\def\lh{\ensuremath{\mathcal{L(H)}}}

\newcommand{\abis}{\stackrel{\lambda}\sim}
\newcommand{\abisa}[1]{\stackrel{#1}\sim}

\newcommand {\nil} {\mbox{\bf{nil}}}
\newcommand {\iif} {\mbox{\bf{if}}}
\newcommand {\then} {\mbox{\bf{then}}}

\newcommand {\true} {\mbox{\bf{true}}}
\newcommand {\false} {\mbox{\bf{false}}}
\renewcommand{\theenumi}{(\arabic{enumi})}
\renewcommand{\labelenumi}{\theenumi}
\newcommand{\tr}{{\rm tr}}
\newcommand{\rto}[1]{\stackrel{#1}\longrightarrow}
\newcommand{\Rto}[1]{\stackrel{#1}\Longrightarrow}

\newcommand{\define}{\stackrel{def}=}

\begin{document}

% Page heads
\markboth{Y. Feng et al.}{Bisimulation for quantum processes}

% Title portion
\title{Bisimulation for quantum processes}
\author{YUAN FENG
\affil{University of Technology, Sydney, Australia, and Tsinghua University, China}
RUNYAO DUAN
\affil{University of Technology, Sydney, Australia, and Tsinghua University, China}
MINGSHENG YING
\affil{University of Technology, Sydney, Australia, and Tsinghua University, China}
}
\begin{abstract}
Quantum cryptographic systems have been commercially available, with a striking advantage over classical systems that their security and ability to detect the presence of eavesdropping are provable based on the principles of quantum mechanics. On the other hand, quantum protocol designers may commit more faults than classical protocol designers since human intuition is poorly adapted to the quantum world. To offer formal techniques for modeling and verification of quantum protocols, several quantum extensions of process algebra have been proposed. 
An important issue in quantum process algebra is to discover a quantum generalization
of bisimulation preserved by various process constructs, in particular,
parallel composition, where one of the major differences between classical
and quantum systems, namely quantum entanglement, is present.
Quite a few versions of bisimulation have been defined for quantum processes in the literature, but in the best case they are only proved to be preserved by parallel composition of purely quantum processes where no classical communication is involved. 

Many quantum cryptographic protocols, however, employ the LOCC (Local Operations and Classical Communication) scheme, where classical communication must be explicitly specified. So, a notion of bisimulation preserved by parallel composition in the circumstance of both classical and quantum communication is crucial for process algebra approach to verification of quantum cryptographic protocols. In this paper we introduce novel notions of strong bisimulation and weak bisimulation for quantum processes, and prove that they are congruent with respect to various process algebra combinators including parallel composition even when both classical and quantum communication are present. We also establish some basic algebraic laws for these bisimulations. In particular, we show the uniqueness of the solutions to recursive equations of quantum processes, which proves useful in verifying complex quantum protocols. To capture the idea that a quantum process approximately implements its specification, and provide techniques and tools for approximate reasoning, a quantified version of strong bisimulation, which defines for each pair of quantum processes a bisimulation-based distance characterizing the extent to which they are strongly bisimilar, is also introduced.  

This is an extended complete version of a paper that was presented at POPL 2011.
\end{abstract}



\begin{bottomstuff}
This work was supported by Australian ARC grants
DP110103473, DP130102764, and FT100100218.
Authors' addresses: 
Center of Quantum Computation \& Intelligent Systems (QCIS), Faculty of Information Technology, University
of Technology, Sydney, City Campus, 15 Broadway, Ultimo, NSW 2007, Australia, and State Key Laboratory of Intelligent Technology
and Systems, Tsinghua National Laboratory for Information Science and Technology, Department
of Computer Science and Technology, Tsinghua University, Beijing 100084, China.
Email addresses:
Y. Feng, Yuan.Feng@uts.edu.au; R. Duan, Runyao.Duan@uts.edu.au; M. Ying,
Mingsheng.Ying@uts.edu.au.
\end{bottomstuff}

\maketitle

\section{Introduction}

Quantum computing offers the potential of considerable speedup over classical computing for some important problems
such as prime factoring \cite{Sh94} and unsorted database search \cite{Gr97}. However, functional quantum computers
which can harness this potential in dealing with practical applications  are extremely difficult to
implement. On the other hand,  quantum cryptography, of which the security and ability to detect
the presence of eavesdropping are provable based on the principles of quantum mechanics,
has been developed so rapidly that quantum
cryptographic systems are already commercially available by a number of companies such as Id Quantique, Cerberis, MagiQ Technologies, SmartQuantum, and NEC. 

As is well known, it is very difficult to guarantee the correctness of classical communication protocols
at the design stage, and some simple protocols were finally found to have fundamental flaws.
Since human intuition is poorly adapted to the quantum world, quantum protocol designers may commit more faults than classical
protocol designers, especially when more and more complicated quantum protocols can be
implemented by future physical technology.
With the purpose of cloning the success classical process algebras achieved in analyzing and verifying
classical communication protocols and even distributed computing, various quantum process algebras have been proposed
independently by several research groups. Jorrand and
Lalire \cite{JL04} defined a language QPAlg (Quantum
Process Algebra) by extending a classical CCS-like process algebra.
A branching bisimulation which identifies quantum processes associated
with graphs having the same branching structure
was also presented \cite{La06}. The bisimulation is,
however, not congruent: it is not preserved by parallel composition.
Gay and Nagarajan \cite{GN05} defined a language CQP
(Communicating Quantum Processes), which combines the communication
primitives of pi-calculus \cite{MP92} with primitives for
unitary transformations and measurements. One
distinctive feature of CQP is a type system which guarantees the
physical realizability of quantum processes. However, no notion of equivalence
between processes was presented.

Authors of the current paper proposed a model named qCCS \cite{FDJY07} for quantum communicating systems by adding quantum input/output and quantum operation/measurement primitives to classical value-passing CCS \cite{He91,HI93}.
The semantics of quantum input and output was carefully designed to describe the communication of quantum systems
which have been entangled with other systems. A bisimulation was defined
for finite processes, and a simplified version of congruence property was proved, in which
parallel composition is only permitted when the participating processes
are free of quantum input, or free of quantum operations and measurements.
In \cite{YFDJ09} the same authors studied a purely quantum version of qCCS where no classical data is explicitly involved, 
aiming at providing a suitable framework to observe the interaction of computation and communication in quantum systems.
A strong bisimulation was defined for this purely quantum qCCS and shown to be fully preserved by parallel composition.
However, it is worth noting that the bisimulation proposed in
\cite{YFDJ09} cannot be directly extended to general qCCS where classical data as well as 
probabilistic behaviors are included. 

In this paper, we combine the two models proposed in \cite{FDJY07} and \cite{YFDJ09} together to involve both classical data and quantum data. This general model, which we still call qCCS for coherence, accommodates all classical process constructors (especially recursive definitions) as well as quantum primitives. As a consequence, both sequential and distributed quantum computing, quantum communication protocols, and quantum cryptographic systems can be formally modeled and rigorously analyzed in the framework of qCCS.
We also design strong/weak bisimulations and approximate strong bisimulation for quantum processes, all turning out to be congruenct with respect to various process constructors of qCCS. These bisimulations have several distinctive features compared with those proposed in the literature: Firstly, the bisimulations in this paper take local quantum operations into account in a weak manner, but at the same time fit well with recursive definitions. Lalire's bisimulation cannot distinguish different operations on a quantum system which will never be output: quantum states are only compared when they are input or output. Bisimulation defined in \cite{FDJY07} works well only for finite processes since quantum states are required to be compared after all the actions have been performed. Note that no state comparison is needed in \cite{YFDJ09} since all local quantum operations are regarded as visible actions, and the resulted bisimulation is a very strong one -- it distinguishes two different sequences of local operations even when they have the same effect as a whole. Secondly, entanglement between the input/output system and the remaining systems is fully considered in our definition of bisimulations. Bisimulation presented in \cite{La06} totally ignores this correlation by only considering the reduced state of the input/output system. In \cite{FDJY07} this consideration is implicitly made by the state comparison after the processes terminating. Again, it does not work for infinite processes. Finally, but most importantly, the strong bisimilarity and the equivalence derived from the weak bisimulation are both congruence, making them suitable for equational reasoning in verifying quantum communication and cryptographic systems.
Lalire's bisimulation is not preserved by parallel composition. The bisimulation in \cite{FDJY07}
is not preserved by restriction, and whether it is preserved by parallel composition still remains open, although the positive answer is affirmed in two special cases. The strong bisimulation proposed in \cite{YFDJ09} is indeed a congruence. However,
since no classical data is involved in that model, many important quantum communication protocols such as superdense coding and teleportation cannot be described. This restricts the scope of its application.

This paper is an extension and completion of our primary results reported at POPL~\cite{FDY11}.
The main difference is that in the current paper (1) a section on strong bisimulation is added where internal actions are treated in the same way as visible actions; (2) a notion of approximate strong bisimulation is introduced to characterize the extent to which two quantum processes are bisimilar; and (3) the proofs of the main results are presented, wheras they were omitted in~\cite{FDY11} because of the limitation of space. The rest of the paper is organized as follows. In Section 2, we review some
basic notions from linear algebra and quantum mechanics. The syntax and operational semantics of qCCS 
are presented in Section 3. To illustrate the expressiveness of qCCS, we describe with it the well-known quantum superdense coding and teleportation protocols. 
We also show how to encode quantum unitary gates and measurement 
gates, which are two basic elements of quantum circuits, by qCCS.
Section~4 defines the notion of strong bisimulation for
configurations as well as quantum processes. Various properties such as congruence property, monoid laws, static laws, the expansion law, as well as uniqueness of solutions of process equations, are also examined. 
In Section 5, a notion of approximate strong bisimulation is proposed and its corresponding metric between quantum processes defined. It is proved that the approximate strong bisimulation is also congruent and the corresponding metric non-expansive with respect to all process constructors in qCCS.
Section~6 is devoted to proposing a weak bisimulation, and an equivalence relation based on the weak bisimularity is also defined and proved to be fully preserved by all
process constructors of qCCS. The validity of examples in Section 3 is proved by using the notion of weak bisimilarity defined in this section. We outline the main results in Section 7 and point out some problems for further study. In particular, we discuss the difficulty of defining an approximate weak bisimulation for quantum processes. 

\section{Preliminaries}
For convenience of the reader, we briefly recall some basic notions
from linear algebra and quantum theory which are needed in this paper. 
For more details, we refer to \cite{NC00}.

\subsection{Basic linear algebra}
An {\it inner product space} $\h$ is a vector space equipped with an inner
product function $$\langle\cdot|\cdot\rangle:\h\times \h\rightarrow \mathbf{C}$$
such that 
\begin{enumerate}
\item
$\langle\psi|\psi\rangle\geq 0$ for any $|\psi\>\in\h$, with
equality if and only if $|\psi\rangle =0$;
\item
$\langle\phi|\psi\rangle=\langle\psi|\phi\rangle^{\ast}$;
\item
$\langle\phi|\sum_i c_i|\psi_i\rangle=
\sum_i c_i\langle\phi|\psi_i\rangle$,
\end{enumerate}
where $\mathbf{C}$ is the set of complex numbers, and for each
$c\in \mathbf{C}$, $c^{\ast}$ stands for the complex
conjugate of $c$. 
Furthermore, if $\h$ is also a complete metric space with respect to the distance function induced by the inner product, then
it is called a {\it Hilbert space}.
For any vector $|\psi\rangle\in\h$, its
length $|||\psi\rangle||$ is defined to be
$\sqrt{\langle\psi|\psi\rangle}$, and it is said to be {\it normalized} if
$|||\psi\rangle||=1$. Two vectors $|\psi\>$ and $|\phi\>$ are
{\it orthogonal} if $\<\psi|\phi\>=0$. An {\it orthonormal basis} of a Hilbert
space $\h$ is a basis $\{|i\rangle\}$ where each $|i\>$ is
normalized and any pair of them are orthogonal.

Let $\lh$ be the set of linear operators on $\h$.  For any $A\in
\lh$, $A$ is {\it Hermitian} if $A^\dag=A$ where
$A^\dag$ is the adjoint operator of $A$ such that
$\<\psi|A^\dag|\phi\>=\<\phi|A|\psi\>^*$ for any
$|\psi\>,|\phi\>\in\h$. The fundamental {\it spectral theorem} states that
the set of all normalized eigenvectors of a Hermitian operator in
$\lh$ constitutes an orthonormal basis for $\h$. That is, there exists
a so-called spectral decomposition for each Hermitian $A$ such that
$$A=\sum_i\lambda_i |i\>\<i|=\sum_{\lambda_{i}\in spec(A)}\lambda_i E_i$$
where the set $\{|i\>\}$ constitute an orthonormal basis of $\h$, $spec(A)$ denotes the set of
eigenvalues of $A$,
and $E_i$ is the projector to
the corresponding eigenspace of $\lambda_i$.
A linear operator $A\in \lh$ is {\it unitary} if $A^\dag A=A A^\dag=I_\h$ where $I_\h$ is the
identity operator on $\h$. In  this paper, we will
use some well-known unitary operators listed as follows: the quantum control-not
operator performed on two qubits with the matrix representation 
$$CN=\left(%
\begin{array}{cccc}
  1 & 0 & 0 & 0 \\
  0 & 1 & 0 & 0 \\
  0 & 0 & 0 & 1 \\
  0 & 0 & 1 & 0
\end{array}%
\right)$$ under the computational basis, and the 1-qubit Hadamard operator $H$ and Pauli operators
$\sigma^0,\sigma^1,\sigma^2,\sigma^3$ defined respectively as
\[
H=\frac{1}{\sqrt{2}}\left(%
\begin{array}{cc}
  1 & 1 \\
  1 & -1 \\
\end{array}%
\right),\ \  \sigma^0=I=\left(%
\begin{array}{cc}
  1 & 0 \\
  0 & 1 \\
\end{array}%
\right),
\]

\[
\sigma^1=\left(%
\begin{array}{cc}
  0 & 1 \\
  1 & 0 \\
\end{array}%
\right),\ \sigma^2=\left(%
\begin{array}{cc}
  1 & 0 \\
  0 & -1 \\
\end{array}%
\right),\ \sigma^3=\left(%
\begin{array}{cc}
  0 & -i \\
  i & 0 \\
\end{array}%
\right).
\]

The {\it  trace} of $A\in\lh$ is defined as $\tr(A)=\sum_i \<i|A|i\>$ for some
given orthonormal basis $\{|i\>\}$ of $\h$. It is worth noting that
trace function is actually independent of the orthonormal basis
selected. It is also easy to check that trace function is linear and
$\tr(AB)=\tr(BA)$ for any operators $A,B\in \lh$.

Let $\h_1$ and $\h_2$ be two Hilbert spaces. Their {\it tensor product} $\h_1\otimes \h_2$ is
defined as a vector space consisting of
linear combinations of the vectors
$|\psi_1\psi_2\rangle=|\psi_1\>|\psi_2\rangle =|\psi_1\>\otimes
|\psi_2\>$ with $|\psi_1\rangle\in \h_1$ and $|\psi_2\rangle\in
\h_2$. Here the tensor product of two vectors is defined by a new
vector such that
$$\left(\sum_i \lambda_i |\psi_i\>\right)\otimes
\left(\sum_j\mu_j|\phi_j\>\right)=\sum_{i,j} \lambda_i\mu_j
|\psi_i\>\otimes |\phi_j\>.$$ Then $\h_1\otimes \h_2$ is also a
Hilbert space where the inner product is defined as the following:
for any $|\psi_1\>,|\phi_1\>\in\h_1$ and $|\psi_2\>,|\phi_2\>\in
\h_2$,
$$\<\psi_1\otimes \psi_2|\phi_1\otimes\phi_2\>=\<\psi_1|\phi_1\>_{\h_1}\<
\psi_2|\phi_2\>_{\h_2}$$ where $\<\cdot|\cdot\>_{\h_i}$ is the inner
product of $\h_i$. For any $A_1\in \mathcal{L}(\h_1)$ and $A_2\in
\mathcal{L}(\h_2)$, $A_1\otimes A_2$ is defined as a linear operator
in $\mathcal{L}(\h_1 \otimes \h_2)$ such that for each
$|\psi_1\rangle \in \h_1$ and $|\psi_2\rangle \in \h_2$,
$$(A_1\otimes A_2)|\psi_1\psi_2\rangle = A_1|\psi_1\rangle\otimes
A_2|\psi_2\rangle.$$  The {\it partial trace} of $A\in\mathcal{L}(\h_1
\otimes \h_2)$ with respect to $\h_1$ is defined as
$\tr_{\h_1}(A)=\sum_i \<i|A|i\>$ where $\{|i\>\}$ is an orthonormal
basis of $\h_1$. Similarly, we can define the partial trace of $A$
with respect to $\h_2$. Partial trace functions are also
independent of the orthonormal basis selected.

A linear operator $\e$ on $\lh$ is {\it completely positive} if it maps
positive operators in $\mathcal{L}(\h)$ to positive operators in
$\mathcal{L}(\h)$, and for any auxiliary Hilbert space $\h'$, the
trivially extended operator $\mathcal{I}_{\h'}\otimes \e$ also maps
positive operators in $\mathcal{L(H'\otimes H)}$ to positive
operators in $\mathcal{L(H'\otimes H)}$. Here $\mathcal{I}_{\h'}$ is
the identity operator on $\mathcal{L(H')}$. The elegant and powerful
{\it Kraus representation theorem} \cite{Kr83} of completely positive
operators states that a linear operator $\e$ is completely positive
if and only if there is some set of operators $\{E_i\}$ with appropriate dimension such that
$$
\e(A)=\sum_{i} E_iA E_i^\dag
$$
for any $A\in \lh$. The operators $E_i$ are called Kraus operators
of $\e$. A linear operator is said to be a {\it super-operator} if it is
completely positive and trace-nonincreasing. Here an operator $\e$ is
{\it trace-nonincreasing} if $\tr(\e(A))\leq \tr(A)$ for any positive $A\in \lh$, and it is said to be
{\it trace-preserving} if the equality always holds. Then a super-operator (resp. a 
trace-preserving super-operator) is 
a completely positive operator
with its Kraus operators $E_i$ satisfying $\sum_i E_i^\dag E_i\leq I$ (resp. $\sum_i E_i^\dag E_i= I$).

\subsection{Basic quantum mechanics}

According to von Neumann's formalism of quantum mechanics
\cite{vN55}, an isolated physical system is associated with a
Hilbert space which is called the {\it state space} of the system. A {\it pure state} of a
quantum system is a normalized vector in its state space, and a
{\it mixed state} is represented by a density operator on the state
space. Here a density operator $\rho$ on Hilbert space $\h$ is a
positive linear operator such that $\tr(\rho)= 1$. 
Another
equivalent representation of density operator is probabilistic
ensemble of pure states. In particular, given an ensemble
$\{(p_i,|\psi_i\rangle)\}$ where $p_i \geq 0$, $\sum_{i}p_i=1$,
and $|\psi_i\rangle$ are pure states, then
$\rho=\sum_{i}p_i[|\psi_i\rangle]$ is a density
operator. Here $[|\psi_i\rangle]$ denotes the abbreviation of
$|\psi_i\>\langle\psi_i|$. Conversely, each density operator can be generated by an
ensemble of pure states in this way.  The set of
density operators on $\h$ is defined as
$$\dh=\{\ \rho\in\lh\ :\  \rho\mbox{ is positive and } \tr(\rho)=
\mbox{1}\}.$$ 

The state space of a composite system (for example, a quantum system
consisting of many qubits) is the tensor product of the state spaces
of its components. For a mixed state $\rho$ on $\h_1 \otimes \h_2$,
partial traces of $\rho$ have explicit physical meanings: the
density operators $\tr_{\h_1}\rho$ and $\tr_{\h_2}\rho$ are exactly
the reduced quantum states of $\rho$ on the second and the first
component system, respectively. Note that in general, the state of a
composite system cannot be decomposed into tensor product of the
reduced states on its component systems. A well-known example is the
 2-qubit state
$$|\Psi\>=\frac{1}{\sqrt{2}}(|00\>+|11\>)
$$
which appears repeatedly in our examples of this paper. This kind of state is called {\it entangled state}.
To see the strangeness of entanglement, suppose a measurement $M=
\lambda_0[|0\>]+\lambda_1[|1\>]$ is applied on the first qubit
of $|\Psi\>$ (see the following for the definition of
quantum measurements). Then after the measurement, the second qubit will
definitely collapse into state $|0\>$ or $|1\>$ depending on whether
the outcome $\lambda_0$ or $\lambda_1$ is observed. In other words,
the measurement on the first qubit changes the state of the second
qubit in some way. This is an outstanding feature of quantum mechanics
which has no counterpart in classical world, and is the key to many
quantum information processing tasks  such as teleportation
\cite{BB93} and superdense coding \cite{BW92}.

The evolution of a closed quantum system is described by a unitary
operator on its state space: if the states of the system at times
$t_1$ and $t_2$ are $\rho_1$ and $\rho_2$, respectively, then
$\rho_2=U\rho_1U^{\dag}$ for some unitary operator $U$ which
depends only on $t_1$ and $t_2$. In contrast, the general dynamics which can occur in a physical system is
described by a trace-preserving super-operator on its state space. 
Note that the unitary transformation $U(\rho)=U\rho U^\dag$ is
a trace-preserving super-operator. 

A quantum {\it measurement} is described by a
collection $\{M_m\}$ of measurement operators, where the indices
$m$ refer to the measurement outcomes. It is required that the
measurement operators satisfy the completeness equation
$\sum_{m}M_m^{\dag}M_m=I_\h$. If the system is in state $\rho$, then the probability
that measurement result $m$ occurs is given by
$$p(m)=\tr(M_m^{\dag}M_m\rho),$$ and the state of the post-measurement system
is $M_m\rho M_m^{\dag}/p(m).$ 

A particular case of measurement is {\it projective measurement} which is usually represented by a Hermitian operator.  Let  $M$ be a
Hermitian operator and
\begin{equation}\label{eq:specdec}
M=\sum_{m\in spec(M)}mE_m
\end{equation} 
its spectral decomposition. Obviously, the projectors  $\{E_m:m\in
spec(M)\}$ form a quantum measurement. If the state of a quantum
system is $\rho$, then the probability that result $m$ occurs when
measuring $M$ on the system is $p(m)=\tr(E_m\rho),$ and the
post-measurement state of the system is $E_m\rho E_m/p(m).$
Note that for each outcome $m$, the map $$\e_m(\rho) =
E_m\rho E_m$$
is again a super-operator by Kraus Theorem; it is not
trace-preserving in general.

Let $M$ be a projective measurement with Eq.(\ref{eq:specdec}) its spectral decomposition. We call $M$ non-degenerate if for any $m\in spec(M)$, the corresponding projector $E_{m}$ is 1-dimensional; that is, all eigenvalues of $M$ are non-degenerate. Non-degenerate measurement is obviously a very special case of general quantum measurement. However, when an ancilla system lying at a fixed state is provided, non-degenerate measurements together with unitary operators are sufficient to implement general measurements. For convenience of the readers, we elaborate the simulation process here. Suppose we are given a quantum system, which we call the principle system in the following, and want to perform a measurement $\{M_{m}\}$ on it. To do this, we introduce an ancilla system having an orthonormal basis $\{|m\>\}$ in one-to-one correspondence with the possible outcomes of the measurement. Let the fixed state of the ancilla system be $|0\>$. We define an operator $U$ such that for any $|\psi\>$,
$$U|\psi\>|0\>  = \sum_{m} M_{m}|\psi\>|m\>.$$
It is direct to check that $U$ can be extended to a unitary operator which we also denote by $U$, from the completeness equation of $\{M_{m}\}$. Now we perform a non-degenerate projective measurement $M=\sum_{m} m|m\>\<m|$ on the ancilla system. Let $\rho$ be the state of the principle system before measurement, and $\sum_{i}p_{i}|\psi_{i}\>\<\psi_{i}|$ be the spectral decomposition of $\rho$. Then for each $i$, $(I\otimes |m\>\<m|)  U|\psi_{i}\>|0\> =  M_{m}|\psi_{i}\>|m\>$. Thus with probability
$$p(m)=\tr [(I\otimes |m\>\<m|) U[\rho\otimes |0\>\<0|] U^{\dag}] =\sum_{i}p_{i} \<\psi_{i}|M_{m}^{\dag}M_{m}|\psi_{i}\>=\tr(M_{m}^{\dag}M_{m}\rho)$$
the outcome $m$ occurs, and the post-measurement states of the principle-ancilla joint system and the principle system,
when $m$ is observed, are given by
$$\frac{(I\otimes |m\>\<m|) U[\rho\otimes |0\>\<0|] U^{\dag}(I\otimes |m\>\<m|) }{\sqrt{p(m)}}= \frac{M_{m}\rho M_{m}^{\dag}\otimes|m\>\<m|}{\tr(M_{m}^{\dag}M_{m}\rho)}$$
and  $M_{m}\rho M_{m}^{\dag}/\tr(M_{m}^{\dag}M_{m}\rho)$, respectively,
which coincide exactly with the case when the measurement $\{M_{m}\}$ is directly applied on the principle system.

We shall need a notion of distance between quantum states in defining approximate strong
bisimulation between quantum processes. For any positive operator
$A$, if $A = \sum_{\lambda_{i}\in spec(A)}\lambda_{i}E_{i}$ is a spectral decomposition of $A$, then we
define
$$\sqrt{A}= \sum_{\lambda_{i}\in spec(A)}\sqrt{\lambda_{i}}E_{i}.$$Furthermore, for any operator A, we set $|A| =
\sqrt{A^{\dag}A}$.  Then the trace distance of $\rho, \sigma\in \d(\h)$ is defined to be
$$d(\rho, \sigma) = \frac{1}{2}\tr|\rho-\sigma|.$$
Trace distance is one of the most popular
metrics used by the quantum
information community. Here we collect some properties of the trace distance which are useful in this paper.

\begin{theorem}\label{thm:disoptimal}(\cite{NC00}, Theorem 9.1) Let $\rho, \sigma\in \d(\h)$. Then
$$d(\rho, \sigma) = \max_{\{M_{i}\}} d(\{p_{i}\}, \{q_{i}\})$$
where the maximization is over all  quantum measurement $\{M_{i}\}$, and $p_{i} = \tr(\rho M^{\dag}_{i} M_{i})$ and $q_{i} = \tr(\sigma M^{\dag}_{i}  M_{i})$ are the probabilities of obtaining outcome $i$ when the initial states are $\rho$ and $\sigma$, respectively. The trace distance between two probabilistic distributions $\{p_{i}\}$ and $\{q_{i}\}$ is defined as
$d(\{p_{i}\}, \{q_{i}\}) = \frac{1}{2}\sum_{i} |p_{i} - q_{i} |$.
\end{theorem}

\begin{theorem}\label{thm:dissup}(\cite{NC00}, Theorem 9.2) Let $\rho, \sigma\in \d(\h)$, and $\e$ a trace-preserving super-operator on $\h$. Then $d[\e(\rho), \e(\sigma)] \leq d(\rho, \sigma).$
\end{theorem}

The notion of trace distance can be extended to super-operators in
a natural way~\cite{Ki97}. For any super-operators $\e_{1}$ and $\e_{2}$ on $\h$, their diamond
trace distance is defined to be
$$d_{\diamond}(\e_{1}, \e_{2}) = \sup\{d[(\e_{1}\otimes I_{\h'} )(\rho), (\e_{2}\otimes I_{\h'} )(\rho)] : \rho\in \d(\h\otimes \h')\}$$
where $\h'$ ranges over all finite-dimensional Hilbert spaces. The quantity $d_{\diamond}(\e_{1}, \e_{2})$ characterizes
the maximal probability that the outputs of $\e_{1}$ and $\e_{2}$ can be distinguished
for the same input where auxiliary systems are allowed.

\section{Basic definitions of qCCS}
In this section, we give the basic definitions of qCCS which is a combination of those proposed in \cite{FDJY07} and \cite{YFDJ09}, involving classical data as well as quantum data, and all classical process 
constructors (especially the recursive definition) as well as quantum primitives. The reader is referred to \cite{FDJY07} and \cite{YFDJ09} for further examples and explanations of the language.
\subsection{Syntax}
We assume three types of data in qCCS: \qc{Bool} for booleans,  real numbers \qc {Real}
for classical data, and qubits \qc {Qbt} for quantum data. Let $cVar$, ranged over
by $x,y,\dots$, be the set of classical variables, and $qVar$, ranged over by $q,r,\dots$, the set of
 quantum variables. It is assumed that $cVar$ and $qVar$ are both countably infinite.
 We assume a set $Exp$ of classical data expressions over
\texttt{Real}, which includes $cVar$ as a subset and is ranged over by $e,e',\dots$, and a set of boolean-valued expressions $BExp$, ranged over by $b, b',\dots$, with the usual set of boolean operators $\true$, $\false$,
$\neg$, $\wedge$, $\vee$, and $\ra$. In particular, we let $e\bowtie e'$ be a boolean expression for any $e,e'\in Exp$ and $\bowtie \in \{ >, <, \geq, \leq, =\}$.
We further assume that only classical variables can occur free in both data expressions and boolean expressions.
 Let $cChan$
be the set of classical channel names, ranged over by $c,d,\dots$,
and $qChan$ the set of quantum channel names, ranged over by $\qc
c,\qc d,\dots$. Let $Chan=cChan\cup qChan$. A relabeling function
$f$ is a one to one function from $Chan$ to $Chan$ such that
$f(cChan)\subseteq cChan$ and $f(qChan)\subseteq qChan$.

We often abbreviate the indexed set
$\{q_1,\dots,q_n\}$ to $\widetilde{q}$ when $q_1, \dots,q_n$ are
distinct quantum variables and the dimension $n$ is understood. Sometimes we also use $\widetilde{q}$ to denote
the string $q_1\dots q_n$. 
We assume a set of process constant schemes, ranged over by 
$A, B, \dots$. Assigned to each process constant scheme $A$ there is a non-negative 
integer $ar(A)$. If $\widetilde{q}$ is a tuple of distinct quantum variables with
$|\widetilde{q}|=ar(A)$, then $A(\widetilde{q})$ is
called a process constant.

Based on these notations, we now propose the syntax of qCCS as follows.

\begin{definition}\rm(Quantum process)\label{def:qProc}
\rm The set of quantum processes $qProc$ and the free quantum
variable function $qv: qProc\rightarrow 2^{qVar}$ are defined inductively
by the following formation rules:
\begin{enumerate}

\item  $\nil \in qProc$, and $qv(\nil)=\emptyset$;

\item  $A(\widetilde{q})\in qProc$, and $qv(A(\widetilde{q}))=\widetilde{q}$;

\item  $\tau.P \in qProc$, and $qv(\tau.P)=qv(P)$;

\item  $c?x.P \in qProc$, and $qv(c?x.P)=qv(P)$;

\item  $c!e.P \in qProc$, and $qv(c!e.P)=qv(P)$;

\item  $\qc c?q.P \in qProc$, and $qv(\qc c?q.P)=qv(P)-\{q\}$;

\item  If $q\not \in qv(P)$ then $\qc c!q.P \in qProc$, and $qv(\qc
c!q.P)=qv(P)\cup\{q\}$;

\item  $\e[\widetilde{q}].P\in qProc$, and
$qv(\e[\widetilde{q}].P)=qv(P)\cup\widetilde{q}$;

\item  $M[\widetilde{q};x].P\in qProc$, and
$qv(M[\widetilde{q};x].P)=qv(P)\cup\widetilde{q}$;

\item  $P+Q\in qProc$, and $qv(P+Q)=qv(P)\cup qv(Q)$;

\item  If $qv(P)\cap qv(Q)=\emptyset$ then $P\| Q\in qProc$, and
$qv(P\| Q)=qv(P)\cup qv(Q)$;

\item  $P[f]\in qProc$, and $qv(P[f])=qv(P)$;

\item  $P\backslash L\in qProc$, and $qv(P\backslash L)=qv(P)$;

\item $\iif\ b\ \then\ P\in qProc$, and $qv(\iif\ b\ \then\
P)=qv(P)$,

\end{enumerate}
where $P,Q\in qProc$, $c\in cChan$, $x\in cVar$, $\qc c\in qChan$,
$q\in qVar$, $\widetilde{q}\subseteq qVar$, $e\in Exp$, $\tau$ is the silent action,
$A(\widetilde{q})$ is a
process constant, $f$ is a relabeling function, $L\subseteq Chan$,
$b\in BExp$, $\e$ and $M$ are respectively
a trace-preserving super-operator and a non-degenerate projective measurement applying on the Hilbert
space associated with the systems $\widetilde{q}$. Furthermore, for each process constant 
$A(\widetilde{q})$, there is a defining equation 
$$A(\widetilde{q})\define P$$
where $P\in qProc$ with $qv(P)\subseteq \widetilde{q}$. When $\widetilde{q}=\emptyset$, we simply
denote $A(\widetilde{q})$ as $A$.
\end{definition}

For the sake of simplicity, we only consider non-degenerate measurements in this paper. This will not sacrifice
the expressiveness of qCCS since as stated in Section 2, non-degenerate measurements can implement general
quantum measurements with the help of unitary operators which, as special case of trace-preserving super-operators, can also be described in qCCS.

The notion of
free classical variables in quantum processes can be defined in the
usual way with a unique modification that the quantum measurement prefix
$M[\widetilde{q};x]$ has binding power on $x$. A quantum process $P$
is closed if it contains no free classical variables, $i.e.$,
$fv(P)=\emptyset$.

\subsection{Operational semantics}

To present the operational semantics of qCCS, some further notations are necessary. 
For each quantum variable $q\in  qVar$, we assume a 2-dimensional 
Hilbert space $\h_q$ to be the
state space of the $q$-system. For any $S\subseteq qVar$,  we denote
$$\h_{S}=\bigotimes_{q\in S} \h_q.$$
In particular, $\h = \h_{qVar}$ is the state space of the whole environment consisting of
all the quantum variables. Note
that $\h$ is a countably-infinite dimensional Hilbert space.

Suppose $P$ is a closed quantum process. A pair of the form
$\<P,\rho\>$ is called a configuration, where $\rho\in \dh$ is a density operator
on $\h$. The set of configurations is denoted by $Con$. We sometimes let  
$\c,\d,\dots$ range over $Con$ to ease notations.

Let $D(Con)$ be the set of finite-support probability distributions
over $Con$; that is,
\begin{eqnarray*}
D(Con)&=&\{\mu:Con\rightarrow [0,1]\ |\
\mu(\c)>0 \mbox{ for finitely}  \mbox{ many $\c$, and }\sum_{\mu(\c)>0}\mu(\c)= 1\}.
\end{eqnarray*}
For any $\mu\in D(Con)$, we denote by $supp(\mu)$ the support set of
$\mu$, $i.e.$, the set of configurations $\c$ such that $\mu(\c)>0$.
When $\mu$ is a simple distribution such that $supp(\mu)=\{\c\}$ for
some $\c$, we abuse the notation slightly to denote $\mu$ by $\c$.
Sometimes we find it
convenient to denote a distribution $\mu$ by an explicit
form $\mu= \boxplus_{i\in I}p_i\bullet \c_i$ (or $\mu= \boxplus
p_i\bullet \c_i$ when the index set $I$ is understood) where $\c_i$ are distinct configurations,
$supp(\mu)=\{\c_i\ :\ i\in I\}$, and $\mu(\c_i)=p_i$ for each $i\in
I$. 

Given $\mu_1,\dots,\mu_n\in D(Con)$ and $p_1,\dots,p_n\in
[0,1]$, $\sum_i p_i=1$, we define the combined distribution, denoted
by $\sum_{i=1}^n p_i\mu_i$, to be a new distribution $\mu$
such that $supp(\mu) = \bigcup_i supp(\mu_i)$, and for any $\d\in supp(\mu)$, $\mu(\d)=\sum_i p_i \mu_i(\d)$.

It is worth pointing out the difference between the two notations $\boxplus_{i\in I} p_i \bullet \c_i$ and $\sum_{i\in I} p_i \c_i$:
the former is the explicit form of a distribution, so it is required that $p_i>0$ for each $i\in I$, and $\c_i\neq \c_j$ for $i\neq j$;
while the latter is the combined distribution of the simple distributions $\c_i$ with the probability weights $p_i$, so
$p_i$ may be zero for some $i\in I$, and $\c_i$s are not necessarily distinct.

Let $\mu= \boxplus_{i\in I} p_i\bullet \<P_i,\rho_i\>$. We denote by $qv(\mu)$ the free variables of $\mu$; that is,   $qv(\mu)=\bigcup_{i\in I} qv(P_i)$. We write $\tr(\mu)=\sum_{i\in I}p_i\tr(\rho_i)$, and $\e(\mu)=\boxplus_{i\in I}p_i\bullet \<P_i, \e(\rho_i)\>$ when $\e$ is a super-operator.

Let
\begin{eqnarray*}
Act&=&\{\tau\}\cup\{c?v,c!v\ |\ c\in cChan, v\in \qc{Real}\}\cup\{\qc c?r,\qc c!r\ |\ \qc c\in qChan, r\in qVar\}.
\end{eqnarray*}
For each $\alpha\in Act$, we define the bound quantum variables $bv(\alpha)$ of $\alpha$ as 
$bv(\qc c?r) = \{r\}$ and $bv(\alpha)=\emptyset$ if $\alpha$ is not a quantum input.
The  channel names used in action $\alpha$ is denoted by $cn(\alpha)$; 
that is, $cn(c?v) = cn(c!v) = \{c\}$, $cn(\qc c?r) = cn(\qc c!r) = \{\qc c\}$, and $cn(\tau)=\emptyset$.

The semantics of qCCS is
given by the probabilistic labeled transition system
$(Con,Act,\rto{})$, where ${\rto{}}\subseteq Con\times
Act\times D(Con)$ is the smallest relation satisfying the rules
defined in Figs. \ref{fig-rules1} and \ref{fig-rules2} (For brevity, we write $\<P,\rho\>
\rto{\alpha} \mu$ instead of $(\<P,\rho\>,\alpha,\mu)\in
{\rto{}}$. The symmetric forms for Rules \textbf{Inp-Int}, \textbf{Oth-Int}, and \textbf{Sum} are omitted).

The transition relation $\rto{}$ can be lifted to  $D(Con)\times
Act\times D(Con)$ by writing $\mu\rto{\alpha} \nu$ if for any $\c\in supp(\mu)$,
$\c\rto{\alpha} \nu_{\c}$ for some $\nu_{\c}$, and $\nu=\sum_{\c\in supp(\mu)} \mu(\c)\nu_{\c}$.

\begin{figure}[t] {
\[
\begin{array}{rl}
\mbox{\textbf{Tau}}: & \displaystyle \frac{}{
\<\tau.P,\rho\> \rto{\tau} {\<P,\rho\>}}\\
\\
\mbox{\textbf{C-Inp}}: & \displaystyle \frac{}{
\<c?x.P,\rho\> \rto{c?v} {\<P\{v/x\},\rho\>}},\hspace{1em}v\in \qc
{Real}\\
\\
\mbox{\textbf{C-Outp}}:& \displaystyle \frac{}{
\<c!e.P,\rho\> \rto{c!v} {\<P,\rho\>}} ,\hspace{1em}v=\sem{e}\\
\\
\mbox{\textbf{C-Com}}: & \displaystyle \frac{ \<P_1,\rho\>\rto{c?v}
{\<P_1',\rho\>},\hspace{1em} \<P_2,\rho\>\rto{c!v} {\<P_2',\rho\>}}
{ \<P_1\|P_2,\rho\>\rto{\tau}
{\<P_1'\|P_2',\rho\>}}\\
\\
\mbox{\textbf{Q-Inp}}: & \displaystyle \frac{}{ \<\qc c?q.P,\rho\>
\rto{\qc c?r} {\<P\{r/q\},\rho\>}},\hspace{1em}r\not\in qv(\qc c?q.P)\\
\\
\mbox{\textbf{Q-Outp}}: & \displaystyle \frac{}{ \<\qc c!q.P,\rho\>
\rto{\qc c!q} {\<P,\rho\>}}\\
\\
\mbox{\textbf{Q-Com}}: & \displaystyle \frac{ \<P_1,\rho\>\rto{\qc
c?r} {\<P_1',\rho\>},\hspace{1em} \<P_2,\rho\>\rto{\qc c!r}
{\<P_2',\rho\>}}{ \<P_1\|P_2,\rho\>\rto{\tau}
\<P_1'\|P_2',\rho\>}\\
\\
\mbox{\textbf{Oper}}:&
\displaystyle \frac{}{ \<\e[\widetilde{r}].P ,\rho\>
\rto{\tau} {\<P,\e_{\widetilde{r}}(\rho)\>}}\\
\\
\mbox{\textbf{Meas}}: & \displaystyle \frac{}{
\<M[\widetilde{r};x].P ,\rho\> \rto{\tau}\sum_{i\in I}
p_i \<P\{\lambda_i/x\},E^i_{\widetilde{r}}\rho
 E^i_{\widetilde{r}}/p_i\>} \\
\\ &\hspace{1em} \mbox{where $M$ has the spectral decomposition }\\
&\hspace{1em} M=\sum_{i\in I} \lambda_i E^i \mbox{ and } p_i=\tr(E^i_{\widetilde{r}}\rho)
\end{array}
\]}
 \caption{Inference rules for qCCS (Part 1)\label{fig-rules1}}
\end{figure}

\begin{figure}[t] {
\[
\begin{array}{rl}
\mbox{\textbf{Inp-Int}}: & \displaystyle \frac{ \<P_1,\rho\>\rto{\qc
c?r} \<P_1',\rho\>}{ \<P_1\|P_2,\rho\>\rto{\qc c?r}
\<P_1'\|P_2,\rho\>},\hspace{1em}\ r\not\in qv(P_2)
\\ \\
\mbox{\textbf{Oth-Int}}: & \displaystyle \frac{
\<P_1,\rho\>\rto{\alpha} \boxplus_{i\in I}
p_i\bullet\<P_i',\rho_i\>}{ \<P_1\|P_2,\rho\>\rto{\alpha}
\boxplus_{i\in I} p_i\bullet\<P_i'\|P_2,\rho_i\>},\hspace{1em}\ \alpha\neq \qc c?r
\\ \\
\mbox{\textbf{Sum}}: & \displaystyle \frac{ \<P,\rho\>\rto{\alpha}
\mu}{ \<P+Q,\rho\>\rto{\alpha} \mu} \\
\\
\mbox{\textbf{Rel}}: & \displaystyle \frac{
\<P,\rho\>\rto{\alpha}\boxplus p_i\bullet \<P_i,\rho_i\>}{
\<P[f],\rho\>\rto{f(\alpha)} \boxplus p_i\bullet \<P_i[f],\rho_i\>}
\\ \\
\mbox{\textbf{Res}}: &
 \displaystyle \frac{ \<P,\rho\>\rto{\alpha} \boxplus p_i\bullet
\<P_i,\rho_i\>}{ \<P\backslash L,\rho\>\rto{\alpha} \boxplus
p_i\bullet \<P_i\backslash L,\rho_i\>},\hspace{1em}\ cn(\alpha)\not\subseteq L
\\ \\
\mbox{\textbf{Cho}}: & \displaystyle \frac{
\<P,\rho\>\rto{\alpha}\mu}{ \<\iif\ b\ \then\
P,\rho\>\rto{\alpha}\mu},\hspace{1em}\ \sem{b}=\mbox{true} \\
\\
 \mbox{\textbf{Def}}: & \displaystyle \frac{
\<P\{\widetilde{r}/\widetilde{q}\},\rho\>\rto{\alpha}\mu}{
\<A(\widetilde{r}),\rho\>\rto{\alpha}\mu},\hspace{1em}\ A(\widetilde{q})\define P
\end{array}
\]}
 \caption{Inference rules for qCCS (Part 2)\label{fig-rules2}}
\end{figure}

For any $S\subseteq qVar$ we denote by $\overline{S}$ the 
complement set of $S$ in $qVar$.
The following lemmas can be easily observed from the
inference rules defined above. 

\begin{lemma}\label{lem:qvchange}
If $\<P,\rho\>\rto{\alpha}\mu$, then
$qv(\mu)\subseteq qv(P)\cup bv(\alpha)$.
\end{lemma}
\begin{proof} By induction on the inference rules.  
\end{proof}

\begin{lemma}\label{lem:superoperator} If
$\<P,\rho\>\rto{\alpha}\mu$, then
\begin{enumerate}

\item $\tr(\rho)=\tr(\mu)$;

\item there exist a set of trace-preserving
super-operators $\{\e_i : i\in I\}$ and a set of projectors $\{E_{i}: i\in I\}$, both acting on $\h_{qv(P)}$ and $\sum_{i\in I}E_{i} = I$, such that
for any $\sigma\in \mathcal{D(H)}$,
$$\<P,\sigma\>\rto{\alpha}\sum_{i\in I}
q_i^\sigma \<P_i,\e_i(\sigma)\>$$
where $q_i^\sigma=\tr(E_{i} \sigma)$;

\item for any trace-preserving super-operator $\e$ acting on $\h_{\overline{qv(P)}}$,
$\<P, \e(\rho)\>\rto{\alpha} \e(\mu).$
\end{enumerate}
\end{lemma}
\begin{proof} By induction on the inference rules. The only case deserving an explanation is for (2) when the action is caused by a  measurement prefix $M[\widetilde{q};x]$. Since only non-degenerate projective measurements are considered
in qCCS, we can suppose that $M=\sum_{i\in I}\lambda_{i} |\psi_{i}\>\<\psi_{i}|$ for some orthonormal basis $\{|\psi_{i}\}$ in the state space of $\widetilde{q}$. Then from the inference rule \textbf{Meas}, we have $$\<P,\sigma\>\rto{\alpha}\sum_{i\in I}
\tr(|\psi_{i}\>\<\psi_{i}|\sigma) \<P\{\lambda_{i}/x\}, |\psi_{i}\>\<\psi_{i}|_{\widetilde{q}}\otimes \sigma'\>$$ where $\sigma'=\tr_{\widetilde{q}}(\sigma)$.  By letting $\e_{i}$ be the trace-preserving super-operator which sets the quantum systems $\widetilde{q}$ to $|\psi_{i}\>$,  $E_{i}=|\psi_{i}\>\<\psi_{i}|$, and $P_{i}= P\{\lambda_{i}/x\}$, the result follows.  \end{proof}

\subsection{Examples}
To illustrate the expressiveness of qCCS, we give some examples.
\begin{example}\label{exam:sdc}\rm 
Superdense coding \cite{BW92} is a quantum protocol using which two bits of classical information can be faithfully transmitted by sending only one qubit, provided that a maximally entangled state is shared $a$ $priori$ between the sender and the receiver. The protocol goes as follows. Let $|\Psi\>=(|00\>+|11\>)/\sqrt{2}$ be the entangled state shared between the sender Alice and the receiver Bob. Alice applies a Pauli operator on her qubit of $|\Psi\>$ according to which information among the four possibilities she wishes to transmit, and sends her qubit to Bob. With the two qubits in hand, Bob performs a perfect discrimination among the possible states (they are actually the four Bell states $\{\sigma^i\otimes I |\Psi\> : i=0,1,2,3\}$ where $\sigma^i$ are defined in Section 2) 
and retrieves the information Alice has sent.

We now show how to describe the protocol of superdense coding with qCCS.
Let $M$ be a 2-qubit measurement such that $M=\sum_{i=0}^3
i|\tilde{i}\>\<\tilde{i}|$, where $\tilde{i}$ is the binary
expansion of $i$. Let $CN$ be the controlled-not operator and $H$ Hadamard operator. Then the quantum processes participated in
superdense coding protocol can be defined as follows:
\begin{eqnarray*}
Alice_s &=& c?x.\sum_{0\leq i \leq 3}\left(\iif\ x=i\ \then\ \sigma^i[q_1].\qc e!q_1.\nil\right),\\
Bob_s &=& \qc e?q_1.CN[q_1,q_2].H[q_1].M[q_1,q_2; x]. d!x.\nil,\\
Sdc &=& (Alice_s\| Bob_s)\backslash \{\qc e\}.
\end{eqnarray*}

For any $\rho\in \d(\h_{\overline{\{q_1,q_2\}}})$ and $v\in \{0,1,2,3\}$, we have the transitions
\begin{eqnarray}
&&\<Sdc, [|\Psi\>]_{q_1,q_2} \otimes \rho\>\nonumber \\
& \rto{c?v}&\left\<\left(\left(\sum_{0\leq i \leq 3}(\iif\ v=i\ \then\ \sigma^i[q_1].\qc e!q_1.\nil)\right) \|
Bob_s\right)\backslash \{\qc e\}, [|\Psi\>]_{q_1,q_2} \otimes \rho\right\> \nonumber \\
& \rto{\tau}& \<(\qc e!q_1.\nil\|
Bob_s)\backslash \{\qc e\}, \sigma^v_{q_1}([|\Psi\>])\otimes \rho\> \nonumber\\
& \rto{\tau}& \<(\nil\|
CN[q_1,q_2].H[q_1].M[q_1,q_2; x]. d!x.\nil)\backslash \{\qc e\}, \sigma^v_{q_1}([|\Psi\>])\otimes \rho\> \nonumber\\
& \rto{\tau}& \<(\nil\|
H[q_1].M[q_1,q_2; x]. d!x.\nil)\backslash \{\qc e\}, CN_{q_1,q_2}(\sigma^v_{q_1}([|\Psi\>]))\otimes \rho\> \nonumber\\
& \rto{\tau}& \<(\nil\|
M[q_1,q_2; x]. d!x.\nil)\backslash \{\qc e\},  [|\widetilde{v}\>]_{q_1,q_2} \otimes \rho\> \label{eq:sdcmeasure}\\
& \rto{\tau}& \<(\nil\| d!v.\nil)\backslash \{\qc e\}, [|\widetilde{v}\>]_{q_1,q_2}\otimes \rho\>\nonumber\\
& \rto{d!v}& \<(\nil\| \nil)\backslash \{\qc e\}, [|\widetilde{v}\>]_{q_1,q_2}\otimes \rho\>.\nonumber
\end{eqnarray}

Here Eq.(\ref{eq:sdcmeasure}) is calculated as follows:
\begin{eqnarray*}
H_{q_1}(CN_{q_1,q_2}(\sigma^v_{q_1}([|\Psi\>])))  =\left\{%
\begin{array}{l}
H_{q_1}(CN_{q_1,q_2}([\frac{|00\>+|11\>}{\sqrt{2}}])) = [|00\>],  \mbox{ if }v=0 \\
\\
H_{q_1}(CN_{q_1,q_2}([\frac{|10\>+|01\>}{\sqrt{2}}])) = [|01\>],  \mbox{ if }v=1 \\
\\
H_{q_1}(CN_{q_1,q_2}([\frac{|00\>-|11\>}{\sqrt{2}}])) = [|10\>],  \mbox{ if }v=2 \\
\\
H_{q_1}(CN_{q_1,q_2}([\frac{|01\>-|10\>}{\sqrt{2}}])) = [|11\>],  \mbox{ if }v=3. 
\end{array}%
\right.
\end{eqnarray*}
\end{example}
\begin{example}\label{exam:tel}\rm
Quantum teleportation
\cite{BB93} is one of the most important protocols in quantum information theory which
can make use of a maximally entangled state shared between the sender and the
receiver to teleport an unknown quantum state by sending only
classical information. It serves as a key ingredient in many other communication protocols. 
The protocol goes as follows. Let $|\Psi\>_{q_1, q_2}$ be the entanglement state
shared between  the sender Alice and the
receiver Bob, with Alice holding $q_1$ and Bob holding $q_2$.
Let $q$ be the quantum system whose state Alice wants to transmit to Bob.
Alice first applies a quantum control-not operations on $q$ and $q_1$, with $q$ the control qubit and
$q_1$ the target, followed by a Hadamard operator $H$ on $q$. She then measures $q$ and $q_1$ according
to the computational basis, and sends the measurement outcome to Bob. Upon receiving the
classical bits from Alice, Bob applies a corresponding Pauli operator on his qubit $q_2$ to
recover the original state of $q$.

Let $M$, $CN$, $H$, and $\sigma^i,\ i=0,\dots,3$ be
as defined in Example \ref{exam:sdc}. Then the quantum processes participated in
teleportation protocol can be defined as follows:
\begin{eqnarray*}
Alice_t &=& \qc c?q.CN[q,q_1].H[q].M[q,q_1;x].e!x.\nil, \\
Bob_t &=& e?x.\sum_{0\leq i \leq 3}(\iif\ x=i\ \then\ \sigma^i[q_2].\qc d!q_2.\nil),\\
Tel &=& (Alice_t\| Bob_t)\backslash \{e\},
\end{eqnarray*}
For any $\rho\in \d(\h_{\overline{\{q_1,q_2\}}})$, we have
\begin{eqnarray}
&&\<Tel, [|\Psi\>]_{q_1,q_2} \otimes \rho\>\nonumber \\
& \rto{\qc c?r}& \<(CN[r,q_1].H[r].M[r,q_1;x].e!x.\nil\|
Bob_t)\backslash \{e\}, [|\Psi\>]_{q_1,q_2} \otimes \rho\> \nonumber \\
& \rto{\tau}& \<(H[r].M[r,q_1;x].e!x.\nil\|
Bob_t)\backslash \{e\}, CN_{r,q_1}([|\Psi\>]_{q_1,q_2} \otimes \rho)\> \nonumber\\
& \rto{\tau}& \<(M[r,q_1;x].e!x.\nil\|
Bob_t)\backslash \{e\}, \sum_{0\leq j\leq 3} \frac{1}{4} [|\widetilde{j}\>]_{r,q_1} \otimes \sigma^j_{q_2}(\rho)\> \label{eq:telmeasure}\\
& \rto{\tau}& 1/4 \bullet \<(e!0.\nil\| Bob_t)\backslash \{e\}, [|00\>]_{r,q_1} \otimes \rho\> \nonumber\\
&  & \boxplus 1/4 \bullet \<(e!1.\nil\| Bob_t)\backslash \{e\}, [|01\>]_{r,q_1} \otimes \sigma_{q_2}^1(\rho)\>\nonumber\\
&  & \boxplus   1/4 \bullet \<(e!2.\nil\| Bob_t)\backslash \{e\}, [|10\>]_{r,q_1} \otimes \sigma_{q_2}^2(\rho)\> \nonumber\\
&  & \boxplus    1/4 \bullet \<(e!3.\nil\| Bob_t)\backslash \{e\}, [|11\>]_{r,q_1} \otimes \sigma_{q_2}^3(\rho)\>,\nonumber
\end{eqnarray}
and for $0\leq j\leq 3$,
\begin{eqnarray*}
& & \<(e!j.\nil\| Bob_t)\backslash \{e\}, [|\widetilde{j}\>]_{r,q_1} \otimes \sigma^j_{q_2}(\rho)\>\\
& \rto{\tau}& \<(\nil\| \sum_{0\leq i \leq 3}(\iif\ j=i\ \then\ \sigma^i[q_2].\qc d!q_2.\nil))\backslash \{e\}, [|\widetilde{j}\>]_{r,q_1} \otimes \sigma^j_{q_2}(\rho)\>\\
& \rto{\tau}& \<(\nil\| \qc d!q_2.\nil)\backslash \{e\},  [|\widetilde{j}\>]_{r,q_1} \otimes \rho\>\\
& \rto{\qc d!q_2}& \<(\nil\| \nil)\backslash \{e\}, [|\widetilde{j}\>]_{r,q_1} \otimes \rho\>.
\end{eqnarray*}
Here Eq.(\ref{eq:telmeasure}) is calculated as follows.
Notice that any $\rho\in \d(\h_{\overline{\{q_1,q_2\}}})$ can be decomposed as 
$\rho = \sum_{0\leq i\leq 3} \gamma_i [|\psi_i\>]_{r} \otimes \rho_i$ where $|\psi_0\>=|0\>,|\psi_1\>=|1\>,|\psi_2\>=|+\>=(|0\>+|1\>)/\sqrt{2},$ and $|\psi_3\>=|-\>=(|0\>-|1\>)/\sqrt{2}$. Then it is easy to derive that 
\begin{eqnarray*}
H_{r}(CN_{r,q_1}([|\Psi\>]_{q_1,q_2} \otimes \rho)) 
 &=& \frac{\gamma_0}{4} [|000\> + |011\> + |100\> + |111\>]_{r,q_1,q_2} \otimes \rho_0\\
& & +\frac{\gamma_1}{4} [|001\> + |010\>  - |101\>  - |110\>]_{r,q_1,q_2} \otimes \rho_1\\
 && + \frac{\gamma_2}{4} [|00+\> + |01+\> + |10-\>  - |11-\>]_{r,q_1,q_2} \otimes \rho_2\\
& & +\frac{\gamma_3}{4} [|00-\> - |01-\> + |10+\>  +  |11+\>]_{r,q_1,q_2} \otimes \rho_3\\
 &=& \frac{1}{4} [|00\>]_{r,q_1} \otimes \rho+ \frac{1}{4} [|01\>]_{r,q_1} \otimes \sigma_{q_2}^1(\rho)\\
 &&+\frac{1}{4} [|10\>]_{r,q_1} \otimes \sigma_{q_2}^2(\rho)+\frac{1}{4} [|11\>]_{r,q_1} \otimes \sigma_{q_2}^3(\rho).
\end{eqnarray*}
\end{example}

\begin{example}\label{exam:qcircuit}\rm (Encode quantum circuits with qCCS)
Quantum circuits consist of two different types of gates: unitary gates and quantum measurements. We now show how to encode them using qCCS. To ease the notations, we allow quantum channels to input and output multiple qubits. We write the quantum channel $\qc c$ as  $\qc c^n$ if $n$ qubits can be communicated through $\qc c$ simultaneously. In other words, the quantum capacity of $\qc c^n$ is $n$ qubits.
\begin{itemize}
\item Unitary gate. Suppose $U$ is a unitary operator acting on $n$ qubits. Then the unitary gate which implements $U$ can be defined as a process constant $\u(U)$, $qv(\u(U))=\emptyset$, with the defining equation
$$\u(U)\define \qc c^n?\widetilde{q}.U[\widetilde{q}].\qc d^n!\widetilde{q}.\u(U).$$
 We set $ar(\u(U))=n$.

\item Measurement gate. Suppose $M$ is a quantum measurement acting on $n$ qubits. Then the measurement gate which implements $M$ can be defined as 
$$\m(M)\define \qc c^n?\widetilde{q}.M[\widetilde{q};x].e!x.\qc d^n!\widetilde{q}.\m(M).$$
We set $ar(\m(M))=n$.
\end{itemize}
For any $\rho\in \dh$, we have
\begin{eqnarray*}
\<\u(U),\rho\> &\rto{\qc c^n?\widetilde{r}}& \<U[\widetilde{r}].\qc d^n!\widetilde{r}.\u(U),\rho\> \\
&\rto{\tau}& \<\qc d^n!\widetilde{r}.\u(U),U_{\widetilde{r}}\rho U_{\widetilde{r}}^\dag\> \\
&\rto{\qc d^n!\widetilde{r}}& \<\u(U),U_{\widetilde{r}}\rho U_{\widetilde{r}}^\dag\> 
\end{eqnarray*}
and
\begin{eqnarray*}
\<\m(M),\rho\> &\rto{\qc c^n?\widetilde{r}}& \<M[\widetilde{r};x].e!x.\qc d^n!\widetilde{r}.\m(M),\rho\> \\
&\rto{\tau}& \boxplus_{i\in I}p_i\bullet \<e!\lambda_i.\qc d^n!\widetilde{r}.\m(M), E^i_{\widetilde{r}}\rho
 E^i_{\widetilde{r}}/p_i\>
\end{eqnarray*}
where $M=\sum_{i\in I} \lambda_i E^i \mbox{ and } p_i=\tr(E^i_{\widetilde{r}}\rho)/\tr(\rho)$. Now for each $i\in I$,
\begin{eqnarray*}
\<e!\lambda_i.\qc d^n!\widetilde{r}.\m(M), E^i_{\widetilde{r}}\rho
 E^i_{\widetilde{r}}/p_i\> 
&\rto{e!\lambda_i}&\<\qc d^n!\widetilde{r}.\m(M), E^i_{\widetilde{r}}\rho
 E^i_{\widetilde{r}}/p_i\>\\
&\rto{\qc d^n!\widetilde{r}}& \<\m(M), E^i_{\widetilde{r}}\rho
 E^i_{\widetilde{r}}/p_i\>.
\end{eqnarray*}

Suppose $\g_1$ and $\g_2$ are two (unitary or measurement) gates with $ar(\g_1)=ar(\g_2)=n$. The sequential composition of $\g_1$ and $\g_2$ can be defined as
\begin{eqnarray*}
\g_1\circ \g_2 &\define& 
(L_s\|\g_1[\qc e^n/\qc c^n, \qc f^n/\qc d^n]\| \g_2[\qc f^n/\qc c^n,\qc g^n/\qc d^n]\|R_s)\backslash\{c, \qc e^n,\qc f^n,\qc g^n\}
\end{eqnarray*}
where $L_s\define \qc c^n?\widetilde{q}.\qc e^n!\widetilde{q}.c?x.L_s$ and $R_s\define \qc g^n?\widetilde{q}.\qc d^n!\widetilde{q}.c!0.R_s$.

If $ar(\g_1)=m$ and $ar(\g_2)=n$, then the parallel composition of $\g_1$ and $\g_2$ is defined as
\begin{eqnarray*}
\g_1\otimes \g_2 &\define&  (L_p\|\g_1[\qc e_1^m/\qc c^m, \qc f_1^m/\qc d^m]\| \g_2[\qc e_2^n/\qc c^n,\qc f_2^n/\qc d^n]\|R_p)\backslash \{c, \qc e_1^m,\qc f_1^m,\qc e_2^n,\qc f_2^n\}
\end{eqnarray*}
 where $L_p\define \qc c^{m+n}?\widetilde{q}.\qc e_1^m!l(\widetilde{q}).\qc e_2^n!r(\widetilde{q}).c?x.L_p$, $R_p\define \qc f_1^m?\widetilde{r_1}.\qc f_2^n?\widetilde{r_2}.\qc d^{m+n}!(\widetilde{r_1}\widetilde{r_2}).c!0.R_p,$  $l(\widetilde{q})$ denotes the prefix of $\widetilde{q}$ with length $m$ while  $r(\widetilde{q})$ the postfix of $\widetilde{q}$ with length $n$, and $\widetilde{r_1}\widetilde{r_2}$ is the concatenation of $\widetilde{r_1}$ and $\widetilde{r_2}$.
\end{example}

\section{Strong bisimulation between quantum processes}

This section is devoted to a strong bisimulation between quantum processes. Firstly, we need a definition from~\cite{BK99} which lifts a relation on $Con$ to a relation on $D(Con)$.
\begin{definition}\label{def:tmp071310} Let $\r\subseteq Con\times Con$, and $\mu,\nu\in D(Con)$.
A weight function for $(\mu, \nu)$ w.r.t. $\r$ is a function $\delta: supp(\mu)\times supp(\nu) \rightarrow [0,1]$ which satisfies
\begin{enumerate}
\item  For all $\c\in supp(\mu)$ and $\d\in supp(\nu)$,
$$
\sum_{\d\in supp(\nu)} \delta(\c,\d)=\mu(\c), \hspace{1em}\sum_{\c\in supp(\mu)} \delta(\c,\d)=\nu(\d);
$$
\item If $\delta(\c,\d)>0,$ then $(\c,\d)\in \r$.
\end{enumerate}
We write $\mu\r \nu$ if there exists a weight function for $(\mu, \nu)$ w.r.t. $\r$. 
\end{definition}

\begin{lemma}\cite{BK99}\label{lem:weight} Suppose $\mu, \nu, \omega\in D(Con)$, $\r, \r'\subseteq Con\times Con$.
\begin{enumerate}
\item
 $\mu{\r} \nu$ if and only if $\nu{\r^{-1}} \mu$;
 \item If $\mu{\r} \nu$ and $\nu{\r'} \omega$, then $\mu(\r\circ\r') \omega$;
 \item If $\r\subseteq \r'$, then $\mu{\r} \nu$ implies $\mu{\r'} \nu$.
\end{enumerate}
\end{lemma}

The following lemma gives an equivalent characterization of the lifted relation on $D(Con)$ directly from the original one on $Con$, without resorting to a weight function:~\footnote{While completing this paper, we were aware of that the same equivalent characterization was established independently by Yuxin Deng and Wenjie Du in \cite{DD11}. They actually adopted Lemma~\ref{lem:weightdec} as the definition of lifted relations, and treated the weight function approach in Definition~\ref{def:tmp071310} as a property.}

\begin{lemma}\label{lem:weightdec} Let $\mu, \nu\in D(Con)$ and $\r\subseteq Con\times Con$. Then $\mu\r \nu$ if and only if  $\mu = \sum_{i\in I} p_i{\c_i}$ and $\nu = \sum_{i\in I} p_i{\d_i}$ such that $\c_i\r \d_i$ for each $i\in I$.   In particular, if $\c\r \nu$ then $\c\r\d$ for each $\d\in supp(\nu)$.
\end{lemma}
\begin{proof} This is simply a special case of Lemma~\ref{lem:lambda} presented in Section 5. \end{proof}

With the notion of lifted relations, we can define strong bisimulation between configurations as follows.
 
\begin{definition}\label{def:sbisimulation}
A relation $\r\subseteq Con\times Con$ is called a 
strong bisimulation if for any $\<P, \rho\>, \<Q, \sigma\>\in Con$, $\<P, \rho\>\r \<Q, \sigma\>$ implies that $qv(P)=qv(Q)$, $\tr_{qv(P)} (\rho) = \tr_{qv(Q)} (\sigma)$, and
\begin{enumerate}

\item whenever $\<P,\rho\> \rto{\qc c?q} \<P',\rho\>$, then $\<Q,\sigma\>\rto{\qc c?q} \<Q', \sigma\>$ for some $Q'$ such that for any trace-preserving super-operator $\e$ acting on $\h_{\overline{qv(P') - \{q\}}}$,  $\<P', \e(\rho)\>\r\<Q', \e(\sigma)\>$;

\item whenever $\<P,\rho\> \rto{\alpha} \mu$ where $\alpha$ is not a quantum input, then there
exists $\nu$ such that $\<Q,\sigma\>\rto{\alpha}\nu$ and
$\mu\r\nu$;

\item whenever $\<Q,\sigma\> \rto{\qc c?q} \<Q', \sigma\>$, then $\<P,\rho\>\rto{\qc c?q}\<P',\rho\>$ for some $P'$ such that for any trace-preserving super-operator $\e$ acting on $\h_{\overline{qv(Q') - \{q\}}}$, $\<P',\e(\rho)\>\r\<Q', \e(\sigma)\>$;

\item whenever $\<Q,\sigma\> \rto{\alpha} \nu$ where $\alpha$ is not a quantum input, then there
exists $\mu$ such that $\<P,\rho\>\rto{\alpha}\mu$ and
$\mu\r\nu$.

\end{enumerate}
\end{definition}

Then the strong bisimilarity between configurations is the largest strong bisimulation, and strong bisimilarity between processes can be defined by comparing two processes in the same environment.

\begin{definition}
\begin{enumerate}
\item Two quantum configurations $\<P, \rho\>$ and $\<Q, \sigma\>$ are strongly bisimilar, denoted by
$\<P, \rho\>\sim \<Q, \sigma\>$, if there exists a strong bisimulation $\r$ such that
$\<P, \rho\>\r \<Q, \sigma\>$;
\item
Two quantum processes $P$ and $Q$ are strongly bisimilar, denoted by
$P\sim Q$, if for any quantum state $\rho\in \d(\h)$ and any indexed
set $\widetilde{v}$ of classical values, $\< P\{\widetilde{v}/\widetilde{x}\}, \rho\>\sim \<Q\{\widetilde{v}/\widetilde{x}\}, \rho \>.$ Here $\widetilde{x}$ is the set of free classical variables contained in $P$ and $Q$.

\end{enumerate}
\end{definition}

Some design decisions made in Definition \ref{def:sbisimulation} deserve justification and explanation:
\begin{itemize}
\item
Recall that in the definition of bisimulations proposed in \cite{FDJY07}, a clause 
\begin{equation}\label{eq:tmp06301}
\mbox{ If $\<P,\rho\>\not\rto{}$ and $\<Q,\sigma\>\not\rto{}$, then $\rho=\sigma$}
\end{equation}
 is presented to guarantee that the quantum operations applied by $P$ and $Q$, which give rise only to invisible actions, have the same effect. That definition, however, does not fit well with recursive definitions since recursively defined processes will generally never reach a terminating process. 

In Definition \ref{def:sbisimulation}, we solve this problem by requiring instead that 
\begin{equation}\label{eq:tmp06302}
\tr_{qv(P)} (\rho) = \tr_{qv(Q)} (\sigma).
\end{equation}
Obviously, when $\<P,\rho\>\not\rto{}$ and $\<Q,\sigma\>\not\rto{}$, and $P$ and $Q$ do not hold any quantum variables, 
Eqs. (\ref{eq:tmp06301}) and (\ref{eq:tmp06302}) are equivalent. However, Eq.(\ref{eq:tmp06302}) can deal with processes which have infinite behaviors. For example, let
$$A\define \qc c?q.Set_{0}[q].\tau.\qc c!q.A$$
and
$$B\define \qc c?q.M_{0,1}[q;x].\sum_{i=0}^1 (\iif\ x=\lambda_i\ \then\ \sigma^{i}[q].\qc c!q.B)$$
where $Set_{0}$ is the trace-preserving super-operator which sets the target qubit to $|0\>$, and $M_{0,1}$ is the 1-qubit
measurement according to the computational basis; that is, $M_{0,1}= \lambda_0 |0\>\<0| + \lambda_1 |1\>\<1|$. 
Intuitively, $B$ can be regarded as an implementation of $A$, specifying how to set the input qubit to $|0\>$. 
We now show $A\sim B$ indeed holds under our definition of strong bisimulation. Let 
\begin{eqnarray*}
Con_{\rho} & =&\{\<A, \rho\>, \<B, \rho\>\} \\
Con_{q,\rho} & = & \{\<A_{1q}, \rho \>, \<A_{2q}, \rho_0 \>, \<A_{3q}, \rho_0 \>, \<B_{1q}, \rho\>, \<B_{2qj}, \rho_j\>, \<B_{3q}, \rho_0\>: j=0,1\}
\end{eqnarray*}
where $A_{1q}=Set_{0}[q].\tau.\qc c!q.A$, $A_{2q}=\tau.\qc c!q.A$, $A_{3q}=\qc c!q.A$,
\begin{eqnarray*}
B_{1q}&=&M_{0,1}[q;x].\sum_{i=0}^1 (\iif\ x=\lambda_i\ \then\ \sigma^{i}[q].\qc c!q.B),\\
B_{2qj}&=&\sum_{i=0}^1 (\iif\ \lambda_j=\lambda_i\ \then\ \sigma^{i}[q].\qc c!q.B), 
\end{eqnarray*}
$B_{3q}=\qc c!q.B$, and $\rho_j=[|j\>]_q\otimes \tr_q \rho$.
Let $\r\subseteq Con\times Con$ such that $\<P,\sigma\>\r \<Q,\eta\>$ if and only if
there exist $q\in qVar$ and $\rho\in \dh$ such that $\<P,\sigma\>$ and $\<Q,\eta\>$ are simultaneously included in $Con_{\rho}$ or $Con_{q,\rho}.$ 
It is not difficult to prove that $\r$ is a strong bisimulation. Thus $A\sim B$.

\item Furthermore, by replacing Eq.(\ref{eq:tmp06301}) with Eq.(\ref{eq:tmp06302}), the derived bisimilarity will be preserved by
restriction. Take the example in \cite{FDJY07}. Let $U_1$, $U_2$, $V_1$, and $V_2$ be unitary
operators such that $U_2U_1 = V_2V_1$ but $U_1\neq V_1$. Let
$$P = U_1[q].c!0.U_2[q].\nil,\ \ Q = V_1[q].c!0.V_2[q].\nil.$$
Then $P$ and $Q$ are strongly bisimilar but $P \backslash\{c\}$ and $Q\backslash\{c\}$ are not if Eq.(\ref{eq:tmp06301}) is required in the definition. However, in our Definition~\ref{def:sbisimulation},  $P \backslash\{c\}$ and $Q\backslash\{c\}$ are also strongly bisimilar since in Eq.(\ref{eq:tmp06302}) we only need to consider the reduced states on the systems $\overline{qv(P)}=\overline{qv(Q)}$. The ``unfinished" quantum operations, which are blocked by the restriction, are not taken into account when comparing the accompanying quantum states.
   
\item Another question one may ask is that why we require $qv(P)=qv(Q)$ in the definition, which excludes the pair
 $$P= I[q].\nil \ \ \ \mbox{ and }\ \ \ Q= \tau.\nil$$ to be strongly bisimilar. 
The reason is, although $P$ and $Q$ have the same effect (they both do nothing at all) on the environment, they are indeed different under parallel composition. For example, if $q\in qv(R)$, then the process $Q\|R$ is valid while $P\|R$ is not.
 
\item
In clause (1), we require $\<P',\e(\rho)\>\r \<Q', \e(\sigma)\>$ for any trace-preserving super-operator $\e$ acting on $\h_{\overline{qv(P')-\{q\}}}$. The reason for this rather strange requirement is as follows. To check whether two configurations are bisimilar, we have to feed them with all possible inputs. In classical process algebra, this is realized by requiring that the input value is arbitrarily chosen. In quantum process algebra, however, since the state of all environmental systems is fixed for a given configuration, only requiring the arbitrariness of the input system is not sufficient. Note that the state-preparation operation and the swap operation are both special trace-preserving super-operators. Our definition actually allows the possibility of inputing an arbitrary system which lies in an arbitrary state.
Furthermore, this requirement is also essential in proving the congruence property of the derived bisimilarity (See Theorems~\ref{thm:sbcpreserve} and \ref{thm:wbcpreserve} below).

\end{itemize}

The following properties can be directly derived from the definitions and Lemma~\ref{lem:weightdec}.
\begin{theorem}\label{thm:slargestbs}
$\sim$ is a strong bisimulation on $Con$, and it is an equivalence relation.
\end{theorem}

\begin{theorem}\label{thm:sbisimulation}
For any configurations $\<P, \rho\>$ and $\<Q, \sigma\>$, $\<P, \rho\>\sim \<Q, \sigma\>$ if and only if $qv(P)=qv(Q)$, $\tr_{qv(P)} (\rho) = \tr_{qv(Q)} (\sigma)$, and
\begin{enumerate}

\item whenever $\<P,\rho\> \rto{\qc c?q} \<P', \rho\>$, then $\<Q,\sigma\>\rto{\qc c?q} \<Q', \sigma\>$ for some $Q'$ such that for any super-operator $\e$ acting on $\h_{\overline{qv(P') - \{q\}}}$,  $\<P', \e(\rho)\>\r\<Q', \e(\sigma)\>$;

\item whenever $\<P,\rho\> \rto{\alpha} \mu$ where $\alpha$ is not a quantum input, then there
exists $\nu$ such that $\<Q,\sigma\>\rto{\alpha}\nu$ and
$\mu\sim\nu$;

\end{enumerate}
and the symmetric conditions of (1) and (2).
\end{theorem}

The strong bisimilarity for configurations is preserved by all $static$ constructors and the summation.

\begin{theorem}\label{thm:sbcpreserve} If $\<P,\rho\>\sim \<Q,\sigma\>$ then
\begin{enumerate}
\item
$\<P+ R,\rho\>\sim \<Q+R,\sigma\>$, provided that $\<R,\rho\>\sim \<R,\sigma\>$;
\item
$\<P\| R,\rho\>\sim \<Q\| R,\sigma\>$;
\item
$\<P[f],\rho\>\sim \<Q[f],\sigma\>$;
\item
$\<P\backslash L,\rho\>\sim \<Q\backslash L,\sigma\>$;
\item
$\<\iif\ b\ \then\ P,\rho\>\sim \<\iif\ b\ \then\ Q,\sigma\>$.
\end{enumerate}
\end{theorem}
\begin{proof}
Items (1) and (3)-(5) are easy from Theorem~\ref{thm:sbisimulation}. Item (2) is simpler than Theorem~\ref{thm:wbcpreserve} (1) in Section~6, thus we omit the proof here.
\end{proof}

The strong configuration bisimilarity is not preserved, however, by $dynamic$ constructors such as prefix. A counterexample is as follows.
Let $P=M_{0,1}[q;x].\nil$ where $M_{0,1}=\lambda_0 [|0\>]+\lambda_1 [|1\>]$ is the 1-qubit measurement according to the computational basis, $Q=I[q].\nil$, and $\rho = [|0\>]_q\otimes \sigma$ where
$\sigma\in \d(\h_{\overline{q}})$. Then $\<P,\rho\>\sim \<Q,\rho\>$, but 
 $\<H[q].P,\rho\>\not\sim \<H[q].Q,\rho\>$ where $H$ is the Hadamard operator.

Nevertheless, similar to classical value-passing CCS, strong bisimilarity for quantum processes is preserved by all
the combinators of qCCS.

\begin{theorem}\label{thm:wbpreserve} If $P\sim Q$ then
\begin{enumerate}
\item
$a.P\sim a.Q$, $a\in \{\tau,c?x,c!e,\qc c?q,\qc
c!q,\e[\widetilde{q}],M[\widetilde{q};x]\}$;
\item
$P+ R\sim Q+R$;
\item
$P\| R\sim Q\|R$;
\item
$P[f]\sim Q[f]$;
\item
$P\backslash L\sim Q\backslash L$;
\item
$\iif\ b\ \then\ P \sim \iif\ b\ \then\ Q$.
\end{enumerate}
\end{theorem}
\begin{proof}
Item (1) is easy to check. The rest is direct from Theorem~\ref{thm:sbcpreserve}.
\end{proof}

The monoid laws and the static laws in classical CCS can also be generalized
to qCCS.

\begin{theorem}
For any $P ,Q, R \in qProc$, $K, L\subseteq Chan$, any relabeling functions $f$ and $f'$, and any action prefix $a$, we have 
\begin{enumerate}
\item $P + \nil \sim P$;
\item $P + P \sim P$;
\item $P + Q \sim Q + P$;
\item $P + (Q + R) \sim (P + Q) + R$;
\item $P\|\nil \sim P$;
\item $P\| Q \sim Q\| P$;
\item $P\|(Q\| R)\sim (P\| Q)\| R$;
\item $(a.P)\backslash L \sim
a.P\backslash L$,   if $cn(a)\not\subseteq L $
\item $(a.P)[f] \sim f(a).P[f]$;
\item $(P+Q)\backslash L \sim P\backslash L + Q\backslash L$;
\item $(P+Q)[f] \sim P[f] + Q[f]$;
\item $P\backslash L \sim P$ if $cn(P) \cap L = \emptyset$, where $cn(P)$ is the set of free channel names used in
$P$;
\item $(P\backslash K)\backslash L \sim P\backslash(K \cup L)$;
\item $(P\|Q)\backslash L \sim P\backslash L\|Q\backslash L$, if
$cn(P)\cap cn(Q)\cap L = \emptyset$;
\item
$P[f]\backslash L\sim P\backslash f^{-1}(L)[f]$;
\item
$P[Id]\sim P$ where $Id$ is the identity relabeling function;
\item
 $P[f]\sim P[f']$
if the restrictions of $f$ and $f'$ on $cn(P)$ coincide;
\item
$P[f][f']\sim P[f'\circ f]$;
\item
$(P\| Q)[f]\sim P[f]\|Q[f]$ if the restriction of $f$ on $cn(P)\cup cn(Q)$
is one-to-one.
\end{enumerate}
\end{theorem}
\begin{proof}
Similar to Propositions 4.7 and 4.8 in \cite{Mi89}
\end{proof}

We now establish the expansion law for quantum processes. In the following theorem, we simply write $P\rto{\alpha} P'$ if for any $\rho\in \dh$, $\<P,\rho\>\rto{\alpha}\<P', \rho\>$.

\begin{theorem}\label{thm:expansion}(Expansion Law) 
Let $$P=(P_1[f_1]\|\cdots \|P_n[f_n])\backslash L.$$ Then
\begin{eqnarray*}
P&\sim  & \sum\left\{ f_i(\alpha).(P_1[f_1]\|\cdots \|P_i'[f_i]\|\cdots \|P_n[f_n])\backslash L:  P_i\rto{\alpha} P_i' \mbox{ and } f_i(cn(\alpha))\not\subseteq L \right\}\\
&+& \sum\left\{ f_i(c)?x.(P_1[f_1]\|\cdots \|P_i'[f_i]\|\cdots \|P_n[f_n])\backslash L:  P_i\rto{c?v} P_i'\{v/x\} \mbox{ for any $v$, and } f_i(c)\not\in L \right\}\\
&+ &  \sum\left\{\e[\widetilde{q}].(P_1[f_1]\|\cdots \|P_i'[f_i]\|\cdots \|P_n[f_n])\backslash L: \<P_i, \rho\> \rto{\tau} \<P_i', \e_{\widetilde{q}}(\rho)\> \mbox{ for any }\rho \right\}\\
&+ & \sum\left\{M[\widetilde{q};x].(P_1[f_1]\|\cdots \|P_i'[f_i]\|\cdots \|P_n[f_n])\backslash L: M=\sum_{j\in J} \lambda_j K^{j}\mbox{ and }\right.\\
&&\hspace{7em} \left.\<P_i, \rho\> \rto{\tau} \sum_{j\in J} p_j\<P_i'\{\lambda_j/x\}, K^{j}_{\widetilde{q}}\rho K^{j}_{\widetilde{q}}/p_j\>\mbox{ for any }\rho  \right\}  \\
&+ &  \sum\left\{\tau.(P_1[f_1]\|\cdots \|P_i'[f_i]\|\cdots \|P_j'[f_j]\|\cdots \|P_n[f_n])\backslash L: \right.\\
&& \hspace{4em}P_i \rto{\alpha} P_i', P_j \rto{\beta} P_j', i<j, f_i(cn(\alpha))=f_j(cn(\beta)), \mbox{and}\\
&& \hspace{7em}\mbox{ among $\alpha$ and $\beta$ there is exactly one input and one output}\}
\end{eqnarray*}
provided that there is at least one summand at the right hand side of the above equation. 
\end{theorem}
\begin{proof}
Similar to Proposition 4.9 in \cite{Mi89}. We put the restriction on the number of summands here for the following reason: in general $Q\backslash L \not\sim \nil$ even if all the free channel names used in $Q$ are included in $L$, since $qv(\nil)=\emptyset$ while $qv(Q\backslash L)=qv(Q)$ is normally not empty. 
\end{proof}

We now turn to examine the properties of strong bisimilarity under recursive definitions. To this end, we assume a set of process variable schemes, ranged over by $X, Y, \dots$. 
 Assigned to each process variable scheme $X$ there is a non-negative 
integer $ar(X)$. If $\widetilde{q}$ is an indexed set of distinct quantum variables with
$|\widetilde{q}|=ar(X)$, then $X(\widetilde{q})$ is
called a process variable.

Process expressions may be defined by adding the following clause into Definition
\ref{def:qProc} (and replacing the word ``process" by the phrase ``process expression" and ``$qProc$" by ``$qExp$"):

\begin{enumerate}
\item[(15)] $X(\widetilde{q})\in qExp$, and $qv(X(\widetilde{q}))=\widetilde{q}$
\end{enumerate}

\noindent where $X(\widetilde{q})$ is a process variable.
We use metavariables $E, F, \dots$ to range over process expressions.
Suppose that $E$ is a process expression, and $\{X_i(\widetilde{q_i}) : i \in I\}$ is a family of
process variables. If $\{P_i : i \in I\}$ is a family of processes such that $qv(P_i)\subseteq \widetilde{q_i}$
for all $i$, then we write
$$E\{P_i/X_i(\widetilde{q_i}) : i \in I\}$$
for the process obtained by replacing simultaneously $X_i(\widetilde{q_i})$ in $E$ with $P_i$ for all $i\in I$.

\begin{definition}
Let $E$ and $F$ be process expressions containing at most process
variables $\{X_i(\widetilde{q_i}) : i\in I\}$. Then $E$ and $F$ are strongly bisimilar, denoted by $E\sim F$, if for all family $\{P_i : i\in I\}$ of quantum processes with $qv(P_i)\subseteq \widetilde{q_i}$,
we have $$E\{P_i/X_i(\widetilde{q_i}) : i\in I\}\sim F\{P_i/X_i(\widetilde{q_i}) : i\in I\}.$$
\end{definition}

For simplicity, sometimes we denote $E\{P_i/X_i(\widetilde{q_i}) : i\in I\}$ as $E\{\widetilde{P}/\widetilde{X}\}$ or even $E(\widetilde{P})$ when it does not cause any confusion. The next theorem shows that $\sim$ is also preserved by recursive definitions.

\begin{theorem}\label{thm:seqpreserve}
\begin{enumerate}
\item
If $A(\widetilde{q})\define P$, then $A(\widetilde{q})\sim P$;
\item
Let $\{E_i : i\in I\}$ and  $\{F_i : i\in I\}$ be two families of process expressions containing at most process
variables $\{X_i(\widetilde{q_i}): i\in I\}$, and $E_i\sim F_i$ for each $i\in I$. If $\{A_i(\widetilde{q_i}) : i\in I\}$ and  $\{B_i(\widetilde{q_i}) : i\in I\}$ be two families of process constants such that
\begin{eqnarray*}
A_i(\widetilde{q_i})&\define&  E_i\{A_j(\widetilde{q_j})/X_j(\widetilde{q_j}) : j\in I\}\\
B_i(\widetilde{q_i})&\define&  F_i\{B_j(\widetilde{q_j})/X_j(\widetilde{q_j}) : j\in I\},
\end{eqnarray*} then $A_i(\widetilde{q_i})\sim B_i(\widetilde{q_i})$ for all $i\in I$.
\end{enumerate}
\end{theorem}
\begin{proof}
(1) is obvious, and (2) is similar to Proposition 4.12 in \cite{Mi89}
\end{proof}

Finally, the uniqueness of solutions of equations can be proved for process expressions in qCCS.

\begin{definition}
Given a process variable $X(\widetilde{q})$ and a process expression $E$, we say $X(\widetilde{q})$ is weakly guarded in $E$ if each occurrence of $X(\widetilde{q})$ is within some subexpression $a.F$ of $E$ where $a$ is a prefix.
\end{definition}

We also say that $E$ is weakly guarded if each process variable is weakly guarded in $E$.

\begin{theorem}\label{thm:suniqueness}
Let $\{E_i : i\in I\}$ be a family of process expressions containing at most process
variables $\{X_i(\widetilde{q_i}): i\in I\}$, and each $X_j(\widetilde{q_j})$ is weakly guarded in each $E_i$. Let $\{P_i : i\in I\}$ and $\{Q_i : i\in I\}$ be two families of quantum processes such that $qv(P_i)\cup qv(Q_i) \subseteq \widetilde{q_i}$ for each $i$, and 
\begin{eqnarray*}
P_i&\sim&  E_i\{P_j/X_j(\widetilde{q_j}) : j\in I\}\\
Q_i&\sim&  E_i\{Q_j/X_j(\widetilde{q_j}) : j\in I\},
\end{eqnarray*} then $P_i\sim Q_i$ for all $i\in I$.
\end{theorem}
\begin{proof}
Similar to Proposition 4.14 in \cite{Mi89}
\end{proof}

\section{Approximate strong bisimulation}

In the previous section, only
$exact$ strong bisimulation is presented where two quantum processes are either bisimilar or non-bisimilar. Obviously, such a
bisimulation cannot capture the idea that a quantum process $approximately$ implements its specification. 
To measure the behavioral distance between processes, the notion of approximate bisimulation and the bisimulation distance 
for classical processes were introduced by various authors~\cite{Yi01,Yi02,DGJP04,DCPP06}.
Note that approximation, or imprecision, is especially essential for quantum process algebra since quantum operations constitute a continuum and exact bisimulation is not always practically suitable for their physical implementation. To provide techniques and tools for approximate reasoning, a quantified version of strong bisimulation, which defines for each pair of quantum processes a bisimulation-based distance characterizing the extent to which they are strongly bisimilar, has already been proposed for purely quantum processes in \cite{YFDJ09}. 
In this section, we introduce an approximate variant of strong bisimulation presented in Section 4. To this end, we first present the approximate notion of weight functions defined in Definition~\ref{def:tmp071310}.

\begin{definition}\label{def:lambdaweight} Let $\r$ be a relation on $Con$, and $\mu,\nu\in D(Con)$.
A $\lambda$-weight function for $(\mu, \nu)$ w.r.t. $\r$ is a function $\delta: supp(\mu)\times supp(\nu) \rightarrow [0,1]$ which satisfies
\begin{enumerate}
\item  For any $\c\in supp(\mu)$ and $\d\in supp(\nu)$,
$$\sum_{\d\in supp(\nu)} \delta(\c,\d)\leq \mu(\c),\ \ \sum_{\c\in supp(\mu)} \delta(\c,\d)\leq \nu(\d);$$
\item  $$\sum_{\c\in supp(\mu)}\sum_{\d\in supp(\nu)} \delta(\c,\d) \geq 1-\lambda;$$
\item If $\delta(\c,\d)>0,$ then $(\c,\d)\in \r$.
\end{enumerate}
We write $\mu\r_{\lambda} \nu$ if there exists a $\lambda$-weight function for $(\mu, \nu)$ w.r.t. $\r$. 
\end{definition}

Similar to Lemma~\ref{lem:weightdec}, we have
\begin{lemma}\label{lem:lambda} Let $\mu, \nu\in D(Con)$. Then $\mu\r_{\lambda} \nu$ if and only if  $\mu = \sum_{i\in I} p_i{\c_i}$ and $\nu = \sum_{i\in I} p_i{\d_i}$ such that $$\sum_{i\in I} \{| p_i : \c_i\r \d_i |\} \geq 1-\lambda.$$  
In particular, for any $\c, \d\in Con$ and $\lambda<1$, $\c\r_{\lambda}\d$ if and only if $\c\r\d$.
\end{lemma}
\begin{proof} Let $\mu\r_{\lambda} \nu$, and $\delta$ is a $\lambda$-weight function for $(\mu, \nu)$ w.r.t. $\r$.
For any $\c\in supp(\mu)$, let $\lambda_{\c}= \mu(\c) - \sum_{\d\in supp(\nu)}\delta(\c, \d)$.
Then we have
\begin{eqnarray*}
\mu &=& \sum_{\c\in supp(\mu)}\mu(\c) \c \\
&=& \sum_{\c, \d}\delta(\c,\d) \c +\sum_{\c\in supp(\mu)}\lambda_\c \c.
\end{eqnarray*}
Similarly, we derive $\nu= \sum_{\c, \d}\delta(\c,\d) \d +\sum_{\d\in supp(\nu)}\lambda_\d \d$ where  $\lambda_{\d}= \nu(\d) - \sum_{\c\in supp(\mu)}\delta(\c, \d)$.
Note that $$\sum_{\c\in supp(\mu)}\lambda_\c =\sum_{\d\in supp(\nu)}\lambda_\d = 1-   \sum_{\c, \d}\delta(\c,\d).$$
We can further write
\begin{eqnarray*}
\mu&=&\sum_{\c, \d}\delta(\c,\d) \c +\sum_{\c,\d}\frac{\lambda_\c\lambda_\d}{T} \c\\
\nu&=&\sum_{\c, \d}\delta(\c,\d) \d +\sum_{\c,\d}\frac{\lambda_\c\lambda_\d}{T} \d
\end{eqnarray*}
where $T=1-   \sum_{\c\r\d}\delta(\c,\d)$.
Let $I_{j}=supp(\mu)\times supp(\nu)\times \{j\}$ for $j=0,1$, and $I=I_{0}\cup I_{1}$. Now for any $(\c,\d,j)\in I$, let
$p_{(\c,\d,j)}$ be $\delta(\c,\d)$ if $j=0$, and $\lambda_\c\lambda_\d/T$ if $j=1$. Furthermore, let $\c_{(\c,\d,j)}=\c$ and  $\d_{(\c,\d,j)}=\d.$
Then
\begin{eqnarray*}
\sum_{(\c,\d,j)\in I} \{| p_{(\c,\d,j)} : \c_{(\c,\d,j)}\r\d_{(\c,\d,j)} |\}
&\geq & \sum_{\c\r\d}\delta(\c,\d)\\
&=&\sum_{\c,\d}\delta(\c,\d)
\geq 1- \lambda.
\end{eqnarray*}
 That proves the necessity part.

Conversely, suppose $\mu = \sum_{i\in I} p_i{\c_i}$ and $\nu = \sum_{i\in I} p_i{\d_i}$ where $\sum_{i\in I} \{| p_i : \c_i\r \d_i |\} \geq 1-\lambda.$ 
Let $I_\c = \{i\in I : \c_i =\c\}$ and $I_\d = \{i\in I : \d_i =\d\}$. 
We construct a function $\delta: supp(\mu)\times supp(\nu) \rightarrow [0,1]$ such that 
$$\delta(\c, \d)=\left\{%
\begin{array}{ll}
  \sum \{| p_i : i\in I_\c\cap I_\d |\}  & \mbox{if}\  \c\r\d, \\ \\
  0 & \mbox{otherwise}.
\end{array}%
\right.$$
Obviously, if $\delta(\c,\d)>0$, then $\c\r\d$. Furthermore, for any $\c\in supp(\mu)$,
\begin{eqnarray*}
\sum_{\d\in supp(\nu)} \delta(\c,\d) &=&\sum \{| p_i : i\in I_\c, \mbox{ and } \c_i\r\d_i |\}\\
&\leq &\sum \{| p_i : i\in I_\c |\}= \mu(\c).
\end{eqnarray*}
Similarly, we have $\sum_{\c\in supp(\mu)} \delta(\c,\d) \leq  \nu(\d).$
Finally, we calculate that
\begin{eqnarray*}\sum_{\c\in supp(\mu)}\sum_{\d\in supp(\nu)}  \delta(\c,\d)  &=&\sum_{\c\in supp(\mu)}\sum \{| p_i : i\in I_\c, \mbox{ and } \c_i\r\d_i |\}\\
&=&\sum \{| p_i : \c_i\r\d_i |\}\geq 1- \lambda.
\end{eqnarray*}
Thus $\delta$ is a $\lambda$-weight function for $(\mu, \nu)$ w.r.t. $\r$, and then $\mu\r\nu$.
\end{proof}

The following lemma is an approximation correspondence of Lemma~\ref{lem:weight}.

\begin{lemma}\label{lem:lambdaweight} Suppose $\mu, \nu, \omega\in D(Con)$, $\r, \r'\subseteq Con\times Con$.
\begin{enumerate}
 \item If $\r\subseteq \r'$ and $\lambda \leq \lambda'$, then $\mu\r_{\lambda} \nu$ implies $\mu \r'_{\lambda'} \nu$.
\item
 $\mu\r_{\lambda} \nu$ if and only if $\nu (\r^{-1})_{\lambda} \mu$;
 \item $\mu\r_{\lambda} \nu$ and $\nu \r'_{\lambda'} \omega$, then $\mu (\r\circ\r')_{\lambda + \lambda'} \omega$;
\end{enumerate}
\end{lemma}
\begin{proof} (1) and (2) are direct from Definition \ref{def:lambdaweight} or Lemma \ref{lem:lambda}. For (3), let $\delta$ be a $\lambda$-weight function for $(\mu, \nu)$ w.r.t. $\r$, and $\delta'$ a $\lambda'$-weight function for $(\nu, \omega)$ w.r.t. $\r'$. We construct $\Delta : supp(\mu)\times supp(\omega) \rightarrow [0,1]$ such that for any $\c\in supp(\mu)$ and $\k\in supp(\omega)$,
$$\Delta(\c, \k) = \sum_{\d\in supp(\nu)} \frac{\delta(\c, \d)\delta'(\d, \k)}{\nu(\d)}.$$
It is easy to check that $\sum_{\k\in supp(\omega)} \Delta(\c, \k) \leq \mu(\c)$ and $\sum_{\c\in supp(\mu)} \Delta(\c, \k) \leq \omega(\k)$. Futheremore,  when $\Delta(\c, \k)>0$, then there exists $\d\in supp(\nu)$ such that both $\delta(\c, \d)>0$ and $\delta'(\d, \k)>0$. Thus $\c\r\d$ and $\d\r'\k$, and so $\c(\r\circ\r')\k$. Finally, we calculate
\begin{eqnarray*}
\sum_{\c\in supp(\mu)}\sum_{\k\in supp(\omega)}\Delta(\c, \k) &=&\sum_{\c\in supp(\mu)} \sum_{\d\in supp(\nu)} \frac{\delta(\c, \d)}{\nu(\d)}(\nu(\d)-\lambda_\d)\\
&\geq& 1-\lambda - \sum_{\c\in supp(\mu)} \sum_{\d\in supp(\nu)} \frac{\delta(\c, \d)}{\nu(\d)}\lambda_\d\\
&\geq& 1-\lambda- \sum_{\d\in supp(\nu)} \lambda_\d \geq 1-\lambda- \lambda'
\end{eqnarray*}
where $\lambda_\d = \nu(\d) - \sum_{\k\in supp(\omega)} \delta'(\d, \k)$, and the last inequality is calculated by 
$$\sum_{\d\in supp(\nu)} \lambda_\d = 1 - \sum_{\d\in supp(\nu)} \sum_{\k\in supp(\omega)} \delta'(\d, \k)\leq \lambda'.$$
Thus $\Delta$ is indeed a $\lambda+\lambda'$-weighted function for $(\mu, \omega)$ w.r.t. $\r\circ\r'$.
\end{proof}

With these notions, we can define the approximate strong bisimulation
between configurations as follows.

\begin{definition}\label{def:asbisimulation}
A relation $\r\subseteq Con\times Con$ is called a 
$\lambda$-strong bisimulation if for any $\<P, \rho\>, \<Q, \sigma\>\in Con$, $\<P, \rho\>\r \<Q, \sigma\>$ implies that $qv(P)=qv(Q)$, $d[\tr_{qv(P)} (\rho), \tr_{qv(Q)} (\sigma)]\leq \lambda$, and
\begin{enumerate}

\item whenever $\<P,\rho\> \rto{\qc c?q} \<P', \rho\>$, then $\<Q,\sigma\>\rto{\qc c?q} \<Q', \sigma\>$ for some $Q'$ such that for any trace-preserving super-operator $\e$ acting on $\h_{\overline{qv(P') - \{q\}}}$,  $\<P', \e(\rho)\>\r_{\lambda}\<Q', \e(\sigma)\>$;

\item whenever $\<P,\rho\> \rto{\alpha} \mu$ where $\alpha$ is not a quantum input, then there
exists $\nu$ such that $\<Q,\sigma\>\rto{\alpha}\nu$ and
$\mu\r_{\lambda}\nu$;
\end{enumerate}
and the symmetric conditions of (1) and (2).
\end{definition}

Note that by Lemma~\ref{lem:lambda}, the $\r_{\lambda}$ in clause (1) of Definition~\ref{def:asbisimulation} can actually replaced by $\r$. Obviously, when $\lambda=0$, the above definition exactly coincides with the strong bisimulation defined in Definition~\ref{def:sbisimulation}. 

The approximate strong bisimilarity between configurations and approximate strong bisimilarity between processes can be defined in a
straightforward way.

\begin{definition}\label{def:tmp060411}
\begin{enumerate}
\item Two quantum configurations $\<P, \rho\>$ and $\<Q, \sigma\>$ are $\lambda$-strongly bisimilar, denoted by
$\<P, \rho\>\abis \<Q, \sigma\>$, if there exists a $\lambda$-strong bisimulation $\r$ such that
$\<P, \rho\>\r \<Q, \sigma\>$;
\item
Two quantum processes $P$ and $Q$ are $\lambda$-strongly bisimilar, denoted by
$P\abis Q$, if for any quantum state $\rho\in \d(\h)$ and any indexed
set $\widetilde{v}$ of classical values, $\< P\{\widetilde{v}/\widetilde{x}\}, \rho\>\abis \<Q\{\widetilde{v}/\widetilde{x}\}, \rho \>.$ Here $\widetilde{x}$ is the set of free classical variables contained in $P$ and $Q$.
\item
The strong bisimulation distance between $P$ and $Q$ is defined by
$$D_{sb}(P,Q)=\inf\{ \lambda\geq 0 : P\abis Q\}.$$
When $P\not\abis Q$ for any $\lambda\geq 0$, we simply set $D_{sb}(P,Q)=\infty$. 
\end{enumerate}
\end{definition}

The following lemmas are useful in proving the latter properties of approximate strong bisimilarity.

\begin{lemma}\label{lem:tmp0415}
For any configurations $\<P, \rho\>$ and $\<Q, \sigma\>$, $\<P, \rho\>\abis \<Q, \sigma\>$ if and only if $qv(P)=qv(Q)$, $d[\tr_{qv(P)} (\rho), \tr_{qv(Q)} (\sigma)]
\leq \lambda$, and
\begin{enumerate}

\item whenever $\<P,\rho\> \rto{\qc c?q}  \<P', \rho\>$, then $\<Q,\sigma\>\rto{\qc c?q} \<Q', \sigma\>$ for some $Q'$ such that for any trace-preserving super-operator $\e$ acting on $\h_{\overline{qv(P') - \{q\}}}$,  $\<P', \e(\rho)\>\abis\<Q', \e(\sigma)\>$;

\item whenever $\<P,\rho\> \rto{\alpha} \mu$ where $\alpha$ is not a quantum input, then there
exists $\nu$ such that $\<Q,\sigma\>\rto{\alpha}\nu$ and
$\mu\abis_{\lambda}\nu$;

\end{enumerate}
and the symmetric conditions of (1) and (2).
\end{lemma}
\begin{proof} Easy from the definitions and Lemma~\ref{lem:lambdaweight}(1).
\end{proof}

\begin{lemma}\label{lem:abisplus}
\begin{enumerate}
\item
If $R_{i}$ is a $\lambda_{i}$-strong bisimulation ($i=1,2$), then $\r_{1}\circ \r_{2}$ is a $(\lambda_{1} + \lambda_{2})$-strong bisimulation;
\item If $\<P, \rho\>\abisa{\lambda_{1}} \<Q,\sigma\>$ and $\<Q, \sigma\>\abisa{\lambda_{2}} \<R,\eta\>$, then $\<P, \rho\>\abisa{\lambda_{1}+\lambda_{2}} \<R,\eta\>$;
\item If $P\abisa{\lambda_{1}} Q$ and $Q\abisa{\lambda_{2}} R$, then $P\abisa{\lambda_{1}+\lambda_{2}} R$;
\item $\abisa{\lambda_{1}}{\subseteq}\abisa{\lambda_{2}}$ whenever $\lambda_{1}\leq \lambda_{2}$.
\end{enumerate}
\end{lemma}
\begin{proof} (1) can be deduced easily from Lemma~\ref{lem:lambdaweight}(3). Then (2) follows from (1), and (3) from (2) directly. Finally, (4) is obvious by definition.
\end{proof}

The following theorem states that the infimum in Definition~\ref{def:tmp060411} (3) of strong bisimulation distance can be replaced by minimum; that is, the infimum is achievable.
\begin{theorem}\label{thm:tmp0604}
If $D_{sb}(P,Q)<\infty$, then $P\abisa{D_{sb}(P,Q)} Q.$
\end{theorem}
\begin{proof} Suppose $\lambda=D_{sb}(P,Q)<\infty$. We need only to prove that
\begin{eqnarray*}
\r=\{(\<P,\rho\>, \<Q,\sigma\>)& : &\<P,\rho\> \abisa{\lambda_{i}} \<Q,\sigma\> \mbox{ for some decreasing sequence }\\
&& \lambda_{1}>\lambda_{2}>\cdots > 0, \mbox{ and } \lim_{i\rightarrow \infty} \lambda_{i}=\lambda\}
\end{eqnarray*}
is a $\lambda$-strong bisimulation. For any $\<P,\rho\>\r \<Q,\sigma\>$, since $\<P,\rho\> \abisa{\lambda_{i}} \<Q,\sigma\>$ we have $qv(P)=qv(Q)$, and 
$d(\tr_{qv(P)}\rho, \tr_{qv(Q)}\sigma)\leq \lambda_{i}$ for any $i\geq 1$. Thus $d(\tr_{qv(P)}\rho, \tr_{qv(Q)}\sigma)\leq \lambda$. Furthermore, 
\begin{enumerate}
\item if $\<P,\rho\>\rto{\qc c?q} \<P', \rho\>$, then for any $i\geq 1$, $\<Q,\sigma\>\rto{\qc c?q} \<Q'_{i}, \sigma\>$ such that for any trace-preserving super-operator $\e$ acting on $\h_{\overline{qv(P') - \{q\}}}$,  $\<P', \e(\rho)\>\abisa{\lambda_{i}}\<Q'_{i}, \e(\sigma)\>$. Since by the semantics of qCCS, all configurations are image-finite; that is, the set $$\mathcal{K} = \{\<Q'_{i}, \sigma\> : \<Q,\sigma\>\rto{\qc c?q} \<Q'_{i}, \sigma\>\}$$ is finite,  there exists a $\<Q', \sigma\>\in \mathcal{K}$ and a decreasing subsequence $\{\lambda_{n_{i}}\}$ of $\{\lambda_{i}\}$ such that  for any trace-preserving super-operator $\e$ acting on $\h_{\overline{qv(P') - \{q\}}}$ and for any $i\geq 1$,  $\<P', \e(\rho)\>\abisa{\lambda_{n_{i}}}\<Q', \e(\sigma)\>$. Thus $\<P', \e(\rho)\>\r \<Q', \e(\sigma)\>$.

 \item if $\<P,\rho\> \rto{\alpha} \mu$ where $\alpha$ is not a quantum input, then for any $i\geq 1$,  $\<Q,\sigma\>\rto{\alpha}\nu_{i}$ and
$\mu\abisa{\lambda_{i}}_{\lambda_{i}}\nu_{i}$. Again, since $\<Q,\sigma\>$ is image-finite, there exists a $\nu\in D(Con)$ and a decreasing subsequence $\{\lambda_{n_{i}}\}$ of $\{\lambda_{i}\}$ such that $\<Q,\sigma\>\rto{\alpha}\nu$, and for any $i\geq 1$,  $\mu\abisa{\lambda_{n_{i}}}_{\lambda_{n_{i}}}\nu$. In the following, we show that this indeed implies $\mu\r_{\lambda}\nu$. 

For any $i\geq 1$, let $\delta_{i} : supp(\mu)\times supp(\nu)\rightarrow [0,1]$ be a $\lambda_{n_i}$-weight function for $(\mu, \nu)$ w.r.t. $\abisa{\lambda_{n_{i}}}$. 
Since $\{\delta_i : i\geq 1\}$ can be regarded as a bounded sequence in the Euclidean space $\mathbb{R}^N$ where $N= |supp(\mu)|\cdot|supp(\nu)|$, there exists  a convergent subsequence $\{\delta_{m_i} \}$ of $\{\delta_i\}$. Let $\delta = \lim_{i\rightarrow \infty} \delta_{m_i}$. Obviously, $\delta$ is again a function from $supp(\mu)\times supp(\nu)$ to $[0,1]$.
Suppose $\delta(\c, \d)>0$. Then there exists $N\geq 1$ such that for any $i\geq N$, $\delta_{m_i}(\c,\d)>0$, and so $\c\abisa{\lambda_{n_{m_i}}}\d$. Thus by the definition of $\r$, we have $\c\r\d$. With this, we can easily check that $\delta$ is a $\lambda$-weight function for $(\mu, \nu)$ w.r.t. $\r$. 

 \end{enumerate}
Symmetric results can be shown when $\<Q,\sigma\>$ performs an action. Thus $\r$ is a $\lambda$-strong bisimulation, from which we  derive easily that $P\abisa{\lambda} Q$.
\end{proof}

A direct consequence of the above theorem is that the strong bisimulation distance between two quantum processes vanishes if and only they are strongly bisimilar.
\begin{corollary} For any $P, Q\in qProc$, $P\sim Q$ if and only if $D_{sb}(P,Q)=0$. \end{corollary}
\begin{proof} Direct from Theorem~\ref{thm:tmp0604}, by noting that $\sim{=}\abisa{0}$. \end{proof}

Similar to strong bisimilarity, the approximation strong bisimilarity is also congruent with respect to various process constructors of qCCS. 

\begin{theorem}\label{thm:lambdacong} For any $\lambda\geq 0$, $\abisa{\lambda}$ is a congruent relation on $qProc$. That is, if $P\abisa{\lambda} Q$ then
\begin{enumerate}
\item
$a.P\abisa{\lambda} a.Q$, $a\in \{\tau,c?x,c!e,\qc c?q,\qc
c!q,\e[\widetilde{q}],M[\widetilde{q};x]\}$;
\item
$P+ R\abisa{\lambda} Q+R$;
\item
$P\| R\abisa{\lambda} Q\|R$;
\item
$P[f]\abisa{\lambda} Q[f]$;
\item
$P\backslash L\abisa{\lambda} Q\backslash L$;
\item
$\iif\ b\ \then\ P \abisa{\lambda} \iif\ b\ \then\ Q$.
\end{enumerate}
\end{theorem}

We now show that all the process constructors of qCCS are non-expansive according to the pseudo-metric $D_{sb}$.  To this end, we need a lemma.
\begin{lemma}\label{lem:tmp0414}
$\<P, \rho\>\abis\<P, \sigma\>$ provided that $d(\rho, \sigma)\leq \lambda$.
\end{lemma}
\begin{proof} We need only to show the following relation $$\r = \{(\<P,\rho\>, \<P,\sigma\>) : d(\rho, \sigma)\leq \lambda \}$$
is a $\lambda$-strong bisimulation. Let $\<P,\rho\>\r \<P,\sigma\>$. Then $d[\tr_{qv(P)}\rho,
\tr_{qv(P)}\sigma]\leq d(\rho, \sigma)\leq \lambda$ by Theorem~\ref{thm:dissup}. Furthermore,
\begin{itemize}
\item if $\<P,\rho\>\rto{\qc c?q} \<P', \rho\>$, then $\<P,\sigma\>\rto{\qc c?q} \<P', \sigma\>$. For any trace-preserving super-operator $\e$ acting on $\h_{qv(P')-\{q\}}$, we have $d[\e(\rho), \e(\sigma)]\leq d(\rho, \sigma)\leq \lambda$, again by Theorem~\ref{thm:dissup}. Thus $\<P',\e(\rho)\>\r \<P',\e(\sigma)\>$ by definition.
\item if $\<P,\rho\>\rto{\alpha}\mu$ is caused by a measurement prefix $M[\widetilde{q};x]$ where $M=\sum_{i\in I}\lambda_{i}|\psi_{i}\>\<\psi_{i}|$, then we have $\mu=\sum_{i\in I}
p_{i}\<P\{\lambda_{i}/x\}, \rho_{i}\>$, $p_{i}=\tr(|\psi_{i}\>\<\psi_{i}|\rho)$, $\rho_{i} = |\psi_{i}\>\<\psi_{i}|_{\widetilde{q}}\otimes \tr_{\widetilde{q}}\rho$,
and $$\<P,\sigma\>\rto{\alpha}\nu=\sum_{i\in I}
q_{i} \<P\{\lambda_{i}/x\}, \sigma_{i}\>$$
with $q_{i}=\tr(|\psi_{i}\>\<\psi_{i}|\sigma) $ and $\sigma_{i}=|\psi_{i}\>\<\psi_{i}|_{\widetilde{q}}\otimes \tr_{\widetilde{q}}\sigma$. Let $\delta : supp(\mu)\times supp(\nu)\rightarrow [0,1]$ such that 
$$\delta(\c, \d)=\left\{%
\begin{array}{ll}
  \min\{p_{i}, q_{i}\}\ \  & \mbox{if}\  \c=\<P\{\lambda_i/x\}, \rho_{i}\> \mbox{ and } \d=\<P\{\lambda_i/x\}, \sigma_{i}\>, \\ \\
  0 & \mbox{otherwise}.
\end{array}%
\right.$$
Then for any $\c\in supp(\mu)$ and $\d\in supp(\nu)$,
$$\sum_{\d\in supp(\nu)}\delta(\c,\d) = \sum_{\c\in supp(\mu)}\delta(\c,\d) = \min\{\mu(\c), \nu(\d)\},$$
and
$$\sum_{\c\in supp(\mu)}\sum_{\d\in supp(\nu)}\delta(\c,\d) = \sum_{i\in I}\min\{p_{i}, q_{i}\} \geq 1-\lambda,$$
where the last inequality is from the following argument. 
Note that 
$$2\sum_{i\in I}\min\{p_{i}, q_{i}\} = \sum_{i\in I}p_{i} + \sum_{i\in I}q_{i} - \sum_{i\in I}|p_{i} - q_{i}|.$$
It follows that $d(\{p_{i}\}, \{q_{i}\}) = 1- \sum_{i\in I}\min\{p_{i}, q_{i}\}$. Furthermore, since $\{|\psi_{i}\>\<\psi_{i}| : i\in I\}$ constitute a quantum measurement on $\widetilde{q}$, we have $d(\{p_{i}\}, \{q_{i}\}) \leq d(\rho, \sigma)$ from Theorem~\ref{thm:disoptimal}.

Now we have shown that $\delta$ is a $\lambda$-weight function for $(\mu, \nu)$ w.r.t. $\r$. Then $\mu\r_{\lambda}\nu$ by definition.
\item if $\<P,\rho\>\rto{\alpha} \mu$ where $\alpha$ is not a quantum input and the transition is not caused by a measurement, then $\mu=\<P', \e(\rho)\>$ for some $P'$ and some trace-preserving super-operator $\e$. Then we have $\<P,\sigma\>\rto{\alpha}\<P', \e(\sigma)\>$, and $\<P', \e(\rho)\>\r\<P', \e(\sigma)\>$.
\end{itemize}
Symmetric results can be shown when $\<P,\sigma\>$ performs an action. Thus $\r$ is a $\lambda$-strong bisimulation.
\end{proof}

\begin{theorem}
\begin{enumerate}
\item The strong bisimulation distance $D_{sb}$ is a pseudo-metric on $qProc$;
\item For any processes $P$ and $Q$, we have:
\begin{enumerate}
\item $D_{sb}(\e[\widetilde{q}].P, \f[\widetilde{q}].Q)\leq d_{\diamond}(\e, \f) + D_{sb}(P,Q)$;
\item $D_{sb}(a.P, a.Q)\leq D_{sb}(P,Q)$ where $a\in \{\tau,c?x,c!e,\qc c?q,\qc
c!q,\e[\widetilde{q}],M[\widetilde{q};x]\}$;
\item $D_{sb}(P+R, Q+R) \leq D_{sb}(P,Q)$;
\item $D_{sb}(P\| R, Q\| R) \leq D_{sb}(P,Q)$;
\item $D_{sb}(P[f], Q[f])\leq D_{sb}(P,Q)$;
\item $D_{sb}(P\backslash L, Q\backslash L ) \leq D_{sb}(P,Q)$;
\item $D_{sb}(\iif\ b\ \then\ P, \iif\ b\ \then\ Q)\leq D_{sb}(P,Q)$.
\end{enumerate}
\end{enumerate}
\end{theorem}
\begin{proof} (1) We need only to prove that $D_{sb}$ satisfies the triangle inequality
$$D_{sb}(P,Q) + D_{sb}(Q,R) \geq D_{sb}(P,R).$$
For any $\lambda_{1}>D_{sb}(P,Q)$ and $\lambda_{2}>D_{sb}(Q,R)$, we have $P\abisa{\lambda_{1}}Q$ and $Q\abisa{\lambda_{2}}R$ by definition. Then $P\abisa{\lambda_{1}+\lambda_{2}}R$ from Lemma~\ref{lem:abisplus}(3). So $ D_{sb}(P,R)\leq \lambda_{1}+\lambda_{2}$, and the result holds from the arbitrariness of $\lambda_{1}$ and $\lambda_{2}$.

(2a) The case when $D_{sb}(P,Q)=\infty$ is obvious. Now suppose $D_{sb}(P,Q)<\infty$. For any $\lambda>D_{sb}(P,Q)$, we have $\<P,\sigma\>\abis\<Q,\sigma\>$ for any $\sigma\in \d(\h)$. To prove the result, it suffices to show $\e[\widetilde{q}].P\abisa{\lambda'} \f[\widetilde{q}].Q$ where $\lambda'=\lambda + d_{\diamond}(\e, \f)$. 

We first derive $qv(P)=qv(Q)$ from $\<P,\rho\>\abis\<Q,\rho\>$, and then $qv(\e[\widetilde{q}].P)=qv(\f[\widetilde{q}].Q)$.
For any $\rho\in \d(\h)$, we have 
$\<\e[\widetilde{q}].P,\rho\>\rto{\tau}\<P,\e_{\widetilde{q}}(\rho)\>$ and
$\<\f[\widetilde{q}].Q,\rho\>\rto{\tau}\<Q,\f_{\widetilde{q}}(\rho)\>$. Note that $d[\e_{\widetilde{q}}(\rho), \f_{\widetilde{q}}(\rho)]\leq d_{\diamond}(\e,\f)$. Then
$\<P,\e_{\widetilde{q}}(\rho)\>\abisa{d_{\diamond}(\e, \f)} \<P,\f_{\widetilde{q}}(\rho)\>$ by Lemma~\ref{lem:tmp0414}. Furthermore, we have $\<P,\e_{\widetilde{q}}(\rho)\>\abisa{\lambda'} \<Q,\f_{\widetilde{q}}(\rho)\>$ from Lemma~\ref{lem:abisplus}(2).   
Thus $\<\e[\widetilde{q}].P, \rho\>\abisa{\lambda'} \<\f[\widetilde{q}].Q, \rho\>$ by Lemma~\ref{lem:tmp0415}, and $\e[\widetilde{q}].P\abisa{\lambda'} \f[\widetilde{q}].Q$ from the arbitrariness of $\rho$. 

(2b) - (2g) are direct from Theorem~\ref{thm:lambdacong}. We omit the proofs here. \end{proof}

Note that in classical process algebra, a notion of approximate bisimulation has been proposed 
for deterministic processes from which any action causes at most one probabilistic transition
\cite{GJS90}. This approximate bisimulation, however, does not yield a pseudo-metric for general probabilistic processes, as shown by van Breugel 
\cite{vB10}. The problem is, Giacalone et al.'s bisimulation is pre-assumed to be an equivalence relation, which, in some sense, violates the intuition that approximate bisimulation is not transitive: $P$ approximates $Q$ and $Q$ approximates $R$ do not necessarily imply that $P$ approximates $R$. Our definition in this section, however, only requires an approximate bisimulation to be symmetric, thanks to the method, introduced by~\cite{BK99}, of lifting relations between processes to those between probability distributions. 
Thus we are able to obtain a pseudo-metric for quantum processes.

\section{Weak bisimulation between quantum processes}

It is obvious that the (approximate) strong bisimulations proposed in previous sections are too overdiscriminative 
since even internal actions,  caused by local quantum operations
and (classical or quantum) communication, are required to be perfectly matched by bisimilar quantum processes. 
In this section, we turn to weak
bisimulation, originated from~\cite{BH97}, which abstracts from the internal actions. To do this,
we first extend the transition relation defined in Section~3.

\begin{definition}\rm\label{def:wtran}
We define the relation ${\Rto{}}\subseteq D(Con)\times D(Con)$ as the smallest 
relation satisfying the following conditions:
\begin{enumerate}
\item $\c\Rto{} \c$;
\item  if $\c \rto{\tau} \mu$ and $\mu \Rto{} \nu$, then $\c\Rto{} \nu$;
\item if $\mu=\sum_{i\in I} p_{i} \c_{i}$, and for any $i\in I$,  $\c_{i}\Rto{}\nu_{i}$ for some $\nu_{i}$, then $\mu\Rto{}\sum_{i\in I} p_{i} \nu_{i}$.
\end{enumerate}
\end{definition}

Allowing different transitions with the same weak labels to be combined together is essential for the definition of weak bisimulation for probabilistic processes, as pointed out in \cite{DPP05} and \cite{DGJP02,DGJP10}. That is the reason why we add clause (3) here in Definition~\ref{def:wtran}.

For any $\mu, \nu\in D(Con)$ and $s=\alpha_1\dots\alpha_n\in Act^*$, we say that $\mu$ can evolve into $\nu$ by a weak
$s$-transition, denoted by $\mu\Rto{s}\nu$, if there exist $\mu_1, \dots, \mu_{n+1},
\nu_1, \dots, \nu_{n}\in D(Con)$, such that $\mu\Rto{}\mu_1$, $\mu_{n+1}= \nu$, and for each $i=1,\dots, n$,
$\mu_i\rto{\alpha_i} \nu_i$ and $\nu_i\Rto{}\mu_{i+1}$.
 
Note that $\mu\Rto{}\rto{\alpha}\nu$ and $\mu\Rto{\alpha}\nu$ are different since in the former the last action
of every execution branch from $\mu$ to $\nu$ must be $\alpha$ while in the latter the action $\alpha$ appeared in each branch is not necessarily
the last one.  

The following lemma is a direct consequence of Proposition 6.1 in \cite{DGH+07}.
\begin{lemma}\label{lem:wtransdec}
If $\mu\Rto{s}\nu$, and $\mu=\sum_{i\in I} p_{i} \mu_{i}$ where $p_{i}>0$ for each $i\in I$, then for any $i\in I$,  $\mu_{i}\Rto{s}\nu_{i}$ for some $\nu_{i}$ such that $\nu=\sum_{i\in I} p_{i} \nu_{i}$. Conversely, if for each $i\in I$, $\mu_{i}\Rto{s}\nu_{i}$, then $\mu\Rto{s}\nu$ where $\mu=\sum_{i\in I} p_{i}\mu_{i}$, $\nu= \sum_{i\in I} p_{i} \nu_{i}$, $p_{i}>0$ for each $i\in I$, and $\sum_{i\in I}p_{i}=1$.
\end{lemma}

By Lemma~\ref{lem:wtransdec}, we can show the transitivity of  weak transitions.

\begin{lemma}\label{lem:wtrans} If $\mu\Rto{}\nu$ and $\nu\Rto{}\omega$, then $\mu\Rto{}\omega$.
\end{lemma}
\begin{proof}We prove by induction on the depth of the inference by which the action $\mu\Rto{}\nu$ is inferred, using clauses (1)-(3) in Definition~\ref{def:wtran}:
\begin{itemize}
\item If $\nu=\mu$, then $\mu\Rto{}\omega$ holds trivially. 
\item Suppose $\mu=\c$, $\c\rto{\tau}\mu'$, and $\mu'\Rto{}\nu$.
Then by induction, we derive $\mu'\Rto{}\omega$. Thus $\mu\Rto{}\omega$ by definition. 
\item Suppose $\mu=\sum_{i\in I} p_{i} \c_{i}$, for any $i\in I$,  $\c_{i}\Rto{}\nu_{i}$ for some $\nu_{i}$, and $\nu=\sum_{i\in I} p_{i} \nu_{i}$. Then by Lemma~\ref{lem:wtransdec}, we have $\nu_{i}\Rto{}\omega_{i}$ for some $\omega_{i}$ such that $\omega = \sum_{i\in I} p_{i}\omega_{i}$. Now by induction, $\c_{i}\Rto{}\omega_{i}$, and then $\mu\Rto{}\omega$ by Lemma~\ref{lem:wtransdec}.
\end{itemize}
\end{proof}

To conclude this subsection, we  extend Lemma~\ref{lem:superoperator} to the weak transition case.

\begin{lemma}\label{lem:superoperator2} If
$\<P,\rho\>\Rto{s}\mu$, then
\begin{enumerate}
\item $\tr(\rho)=\tr(\mu)$;

\item there exist a set of trace-preserving
super-operators $\{\e_i : i\in I\}$ and a set of projectors $\{E_{i}: i\in I\}$, both acting on  $\h_{qv(P)\cup bv(s)}$ where
 $bv(\alpha_1\dots\alpha_n)=bv(\alpha_1)\cup \dots \cup bv(\alpha_n)$, $\sum_{i\in I}E_{i} = I$, such that
for any $\sigma\in \mathcal{D(H)}$,
$$\<P,\sigma\>\Rto{s}\sum_{i\in I}
q_i^\sigma \<P_i,\e_i(\sigma)\>$$
where $q_i^\sigma=\tr(E_{i} \sigma)$;

\item for any trace-preserving super-operator $\e$ acting on $\h_{\overline{qv(P)\cup bv(s)}},$ we have
$\<P, \e(\rho)\>\Rto{s}\e(\mu).$
\end{enumerate}
\end{lemma}
\begin{proof} Note that from Lemma \ref{lem:superoperator} (1), if $\nu\rto{\alpha}\mu$ then $\tr(\nu)=\tr(\mu)$. So to prove (1), we need only to show $\tr(\nu)=\tr(\mu)$ provided that $\nu\Rto{}\mu$. We prove by induction on the depth of the inference by which the action $\nu\Rto{}\mu$ is inferred, using clauses (1)-(3) in Definition~\ref{def:wtran}:
\begin{itemize}
\item If
$\nu=\mu$, then $\tr(\nu)=\tr(\mu)$ holds trivially.
\item Suppose 
$\nu=\<P,\rho\>$, $\<P,\rho\> \rto{\tau}\omega$, and $\omega\Rto{}\mu$.  Then we have
$\tr(\mu)=\tr(\omega)=\tr(\rho)$, where the first equation is derived by induction, and the second by Lemma~\ref{lem:superoperator}(1).
\item Suppose $\nu=\sum_{i\in I} p_{i} \c_{i}$, for any $i\in I$,  $\c_{i}\Rto{}\nu_{i}$ for some $\nu_{i}$, and $\mu=\sum_{i\in I} p_{i} \nu_{i}$. Then by induction, $\tr(\nu_{i})=\tr(\c_{i})$. Thus $\tr(\mu)=\sum_{i\in I}p_{i} \tr(\nu_{i}) =\sum_{i\in I}p_{i} \tr(\c_{i}) = \tr(\nu)$. 
\end{itemize}

The proofs of (2) and (3) are more complicated, but the idea is similar. So we omit the detail here. \end{proof}

\subsection{Weak bisimulation}

\begin{definition}\label{def:wbisimulation}
A relation $\r\subseteq Con\times Con$ is called a 
weak bisimulation if for any $\<P, \rho\>, \<Q, \sigma\>\in Con$, $\<P, \rho\>\r \<Q, \sigma\>$ implies that $qv(P)=qv(Q)$, $\tr_{qv(P)} (\rho) = \tr_{qv(Q)} (\sigma)$, and
\begin{enumerate}

\item whenever $\<P,\rho\> \rto{\qc c?q} \mu$, then $\<Q,\sigma\>\Rto{}\rto{\qc c?q} \nu$ for some $\nu$ such that for any trace-preserving super-operator $\e$ acting on $\h_{\overline{qv(\mu) - \{q\}}}$,  $\e(\mu)\r\e(\nu)$;

\item whenever $\<P,\rho\> \rto{\alpha} \mu$ where $\alpha$ is not a quantum input, then there
exists $\nu$ such that $\<Q,\sigma\>\Rto{\widehat{\alpha}}\nu$ and
$\mu\r\nu$;

\item whenever $\<Q,\sigma\> \rto{\qc c?q} \nu$, then $\<P,\rho\>\Rto{}\rto{\qc c?q}\mu$ for some $\mu$ such that for any trace-preserving super-operator $\e$ acting on $\h_{\overline{qv(\nu) - \{q\}}}$, $\e(\mu)\r\e(\nu)$;

\item whenever $\<Q,\sigma\> \rto{\alpha} \nu$ where $\alpha$ is not a quantum input, then there
exists $\mu$ such that $\<P,\rho\>\Rto{\widehat{\alpha}}\mu$ and
$\mu\r\nu$.

\end{enumerate}
\end{definition}

\begin{definition}
\begin{enumerate}
\item Two quantum configurations $\<P, \rho\>$ and $\<Q, \sigma\>$ are weakly bisimilar, denoted by
$\<P, \rho\>\approx \<Q, \sigma\>$, if there exists a weak bisimulation $\r$ such that
$\<P, \rho\>\r \<Q, \sigma\>$;
\item
Two quantum processes $P$ and $Q$ are weakly bisimilar, denoted by
$P\approx Q$, if for any quantum state $\rho\in \d(\h)$ and any indexed
set $\widetilde{v}$ of classical values, $\< P\{\widetilde{v}/\widetilde{x}\}, \rho\>\approx \<Q\{\widetilde{v}/\widetilde{x}\}, \rho \>.$ Here $\widetilde{x}$ is the set of free classical variables contained in $P$ and $Q$.

\end{enumerate}
\end{definition}

To illustrate the power of weak bisimilarity defined above, we revisit the examples presented in Section 3.
\begin{example}\rm (Superdense coding revisited)
This example is devoted to proving rigorously that the protocol presented in Example \ref{exam:sdc} indeed sends two bits of classical information from Alice to Bob by transmitting a qubit, with the help of a maximally entangled state. 
Let $$Sdc_{spec} = c?x.Set_x[q_1,q_2].d!x.\nil$$ be the specification of superdense coding protocol,
where  $$Set_x[q_1,q_2].d!x.\nil=\sum_{i=0}^3 (\iif\ x=i\ \then\ Set_i[q_1,q_2].d!x.\nil),$$
and $Set_{i}$, $i=0,\dots,3$, is the 2-qubit super-operator which sets the target qubits to $|\widetilde{i}\>$; that  is, for any $\rho\in \d(\h)$, $$Set_{i, q,q'}(\rho)=
[| \widetilde{i}\>]_{q,q'} \otimes \tr_{q,q'} (\rho).
$$
We have $Set_x[q_1,q_2]$ in the specification simply for technical reasons: to make $qv(Sdc_{spec})=qv(Sdc)$, and to set $q_1, q_2$ to the required final states. For any $\rho \in  \d(\h_{\overline{\{q_1,q_2\}}})$, and $v\in \{0,1,2,3\}$,
\begin{eqnarray*}
&&\<Sdc_{spec},  [|\Psi\>]_{q_1, q_2} \otimes\rho\>\\
 & \rto{c?v}& \<Set_v[q_1,q_2].d!v.\nil,  [|\Psi\>]_{q_1, q_2} \otimes\rho\> \\
& \rto{\tau}&  \<d!v.\nil,  [|\widetilde{v}\>]_{q_1, q_2} \otimes\rho\>\\ 
& \rto{d!v}&  \<\nil,  [|\widetilde{v}\>]_{q_1, q_2} \otimes\rho\>. 
\end{eqnarray*}
We can easily prove
$$\<Sdc, [|\Psi\>]_{q_1, q_2} \otimes \rho\> \approx \<Sdc_{spec}, [|\Psi\>]_{q_1, q_2} \otimes \rho\>
$$ by checking that  
\begin{eqnarray*}
\r& =&\ \ \{(\<Sdc, \rho_\Psi\>, \<Sdc_{spec}, \rho_\Psi\>)\} \\
&& \cup  \{(\<P, \eta \>, \<Set_v[q_1,q_2].d!v.\nil,  \rho_\Psi\>):  v=0,\dots, 3,\\
&&\ \ \ \ 
\<Sdc, \rho_\Psi\> \Rto{c?v} \<P, \eta \>, \mbox{ and } qv(P)\neq \emptyset\}\\
&& \cup  \{(\<(\nil\|d!v.\nil)\backslash \{\qc e\}, \rho_{\widetilde{v}}\>, \<d!v.\nil, \rho_{\widetilde{v}}\>):  v=0,\dots, 3\}  \\
&&\cup \{ (\<(\nil\|\nil)\backslash \{\qc e\}, \rho_{\widetilde{v}}\>, \<\nil, \rho_{\widetilde{v}}\>):  v=0,\dots, 3\}
\end{eqnarray*}
 is a weak bisimulation, where $\rho_{\psi}=[|\psi\>]_{q_1, q_2} \otimes \rho$. 
\end{example}

Note that  $Sdc\approx Sdc_{spec}$ does not hold in general since superdense coding protocol needs the assistance of a 
maximally entangled state to realize the intended task.

\begin{example}\rm (Teleportation revisited)
This example is devoted to proving rigorously that the protocol presented in Example \ref{exam:tel} indeed teleports any unknown quantum state from Alice to Bob, again with the help of a maximally entangled state. To employ our notion of weak bisimulation, we need to modify the original definition of Alice's protocol in Example \ref{exam:tel} as follows:
$$Alice_t' = \qc c?q.CN[q,q_1].H[q].M[q,q_1;x].Set_{\Psi}[q,q_1].e!x.\nil$$
and $Tel' = (Alice_t'\|Bob_t)\backslash \{e\}$ where $Set_{\Psi}$ is the 2-qubit super-operator which sets the target qubits to $|\Psi\>$.
Let $$Tel_{spec} = \qc c?q.SWAP_{1,3}[q,q_1,q_2].\qc d!q_2.\nil$$ be the specification of teleportation protocol,
where $SWAP_{1,3}$ is a 3-qubit unitary operator which exchanges the states of the first and the third qubits, keeping the second qubit untouched. Again, we involve qubit $q_1$ here just for technical reason: to make $qv(Tel_{spec})=qv(Tel')$.
Then  for any $\rho\in \d(\h_{\overline{\{q_1,q_2\}}})$ and $r\neq q_1,q_2$,
\begin{eqnarray*}
&&\<Tel_{spec},  [|\Psi\>]_{q_1, q_2} \otimes \rho\>\\
&\rto{\qc c?r}& \<SWAP_{1,3}[r,q_1,q_2].\qc d!q_2.\nil,  [|\Psi\>]_{q_1, q_2} \otimes \rho\> \\
& \rto{\tau}& \<\qc d!q_2.\nil,   [|\Psi\>]_{q_1, r} \otimes \rho\>\\
& \rto{\qc d!q_2}& \<\nil,  [|\Psi\>]_{q_1, r} \otimes \rho\>. 
\end{eqnarray*}
We can now prove
$$\<Tel', [|\Psi\>]_{q_1, q_2} \otimes \rho\> \approx \<Tel_{spec}, [|\Psi\>]_{q_1, q_2} \otimes \rho\>
$$ 
by checking that
\begin{eqnarray*}
\r& =&\ \ \ \{(\<Tel', \rho_{\Psi}^{q_1,q_2}\>, \<Tel_{spec}, \rho_{\Psi}^{q_1,q_2}\>)\} \\
&& \cup  \{(\<P, \eta \>, \<SWAP_{1,3}[r,q_1,q_2].\qc d!q_2.\nil, \sigma_{\Psi}^{q_1,q_2}): \\
&&\ \ \ \ \<Tel', \sigma_{\Psi}^{q_1,q_2}\> \Rto{\qc c?r}  \<P, \eta \>,  \sigma\in \d(\h_{\overline{\{q_1,q_2\}}}), qv(P)= \{r,q_1,q_2\}, \mbox{ and } r\neq q_1,q_2\}  \\
&& \cup  \{(\<P, \eta \>, \<\qc d!q_2.\nil, \sigma_{\Psi}^{q_1,r}): \<Tel', \sigma_{\Psi}^{q_1,q_2}\> \Rto{\qc c?r}  \mu \mbox{ with }  \<P, \eta \> \in supp(\mu),\\
&&\ \ \ \  \sigma\in \d(\h_{\overline{\{q_1,r\}}}),  qv(P)= \{q_2\}, \mbox{ and } r\neq q_1,q_2\}  \\
&&\cup \{ (\<(\nil\|\nil)\backslash \{e\}, \sigma_{\Psi}^{q_1,r}\>, \<\nil,\sigma_{\Psi}^{q_1,r}\>) :\sigma\in \d(\h_{\overline{\{q_1,r\}}})\}
\end{eqnarray*}
 is a weak bisimulation, where $\sigma_{\psi}^{q,q'}=[|\psi\>]_{q, q'} \otimes \sigma$. 
\end{example}

Again,  $Tel'\approx Tel_{spec}$ does not hold in general since teleportation protocol is valid only when a maximally entangled state is provided and consumed. 

\begin{example}\label{exam:ugate}\rm (Encode quantum circuits by qCCS, revisited)
Using the notations presented in Example \ref{exam:qcircuit}, we can prove the following properties considering the sequential composition and parallel composition of quantum gates: 
\begin{enumerate}
\item $\u(U)\circ \u(V)  \approx \u(V U)$, provided that $ar(\u(U))=ar(\u(V))$;
\item $\u(U)\circ \m(M)  \approx  \m(U^\dag MU) \circ\u(U)$, provided that $ar(\u(U))=ar(\m(M))$;
\item $\u(U)\otimes \u(V) \approx \u(U\otimes V)$.
\end{enumerate}
The proof is straightforward, and we only take (1) as an example. Suppose $ar(\u(U))=ar(\u(V))=n$. Let 
\begin{eqnarray*}
\r&=&\{(\<\u(U)\circ \u(V), \rho\>, \<\u(VU), \rho\>) : \rho\in \dh\} \\
&\cup&    \{(\<P, \sigma \>, \<Q, \eta\>): \<\u(U)\circ \u(V), \rho\>\Rto{\qc c^n?\widetilde{r}} \<P, \sigma \>\mbox{ and }\\
&&\<\u(VU), \rho\>\Rto{\qc c^n?\widetilde{r}} \<Q, \eta \> \mbox{ where $\widetilde{r}\subseteq qVar$ and } \rho\in \dh\}\\
&\cup&    \{(\<P, \sigma \>, \<Q, \eta\>): \<\u(U)\circ \u(V), \rho\>\Rto{\qc c^n?\widetilde{r},\qc d^n!\widetilde{r}} \<P, \sigma \>\mbox{ and }\\
&&\<\u(VU), \rho\>\Rto{\qc c^n?\widetilde{r},\qc d^n!\widetilde{r}} \<Q, \eta \> \mbox{ where $\widetilde{r}\subseteq qVar$ and } \rho\in \dh\}.
\end{eqnarray*}
It is easy to check that $\r$ is a weak bisimulation. So we have $\<\u(U)\circ \u(V), \rho\>\approx \<\u(VU), \rho\>$ for all $\rho\in \dh$ and then $\u(U)\circ \u(V)  \approx \u(V U)$.
\end{example}

To conclude this subsection, we prove some properties of weak bisimilarity which are useful in the rest of this paper.

\begin{lemma}\label{lem:exttrans}
Let $\r$ be a weak bisimulation, and $\mu\r\nu$. 
\begin{enumerate}
\item
If $\mu\Rto{}\mu'$, then there exists $\nu'$ such that $\nu\Rto{}\nu'$ and $\mu'\r\nu'$;
\item
If $\mu\rto{\qc c?q}\mu'$, then there exists $\nu'$ such that $\nu\Rto{}\rto{\qc c?q}\nu'$ and for any trace-preserving super-operator $\e$ acting on $\h_{\overline{qv(\mu') - \{q\}}}$,  $\e(\mu')\r\e(\nu')$;
\item
If $\mu\rto{\alpha}\mu'$, then there exists $\nu'$ such that $\nu\Rto{\hat{\alpha}}\nu'$ and $\mu'\r\nu'$;
\end{enumerate}
\end{lemma}
\begin{proof}  Easy from Lemmas~\ref{lem:wtransdec} and \ref{lem:wtrans}.
\end{proof}

\begin{theorem}\label{thm:largestbs}
$\approx$ is a  weak bisimulation on $Con$, and it is an equivalence relation.
\end{theorem}
\begin{proof}  Suppose each $\r_i$, $i=1,2,\dots$, is a weak bisimulation on $Con$. From Lemmas~\ref{lem:weight} and \ref{lem:exttrans}, we can prove that 
the following relations are all weak bisimulations:
\[
\begin{tabular}{ll}
(1) $Id_{Con}$&\hspace{3em}(2) $\r_i^{-1}$\\ 
(3) $\r_1\circ \r_2$&\hspace{3em}(4) $\bigcup_i \r_i$.
\end{tabular}
\]
Then the result follows. \end{proof}

The following lemma gives a recursive characterization of weak bisimilarity
between configurations. 

\begin{theorem}\label{thm:bisimulation}
For any configurations $\<P, \rho\>$ and $\<Q, \sigma\>$, $\<P, \rho\>\approx \<Q, \sigma\>$ if and only if $qv(P)=qv(Q)$, $\tr_{qv(P)} (\rho) = \tr_{qv(Q)} (\sigma)$, and
\begin{enumerate}

\item whenever $\<P,\rho\> \rto{\qc c?q} \mu$, then $\<Q,\sigma\>\Rto{}\rto{\qc c?q} \nu$ for some $\nu$ such that for any trace-preserving super-operator $\e$ acting on $\h_{\overline{qv(\mu) - \{q\}}}$,  $\e(\mu)\approx\e(\nu)$;

\item whenever $\<P,\rho\> \rto{\alpha} \mu$ where $\alpha$ is not a quantum input, then there
exists $\nu$ such that $\<Q,\sigma\>\Rto{\widehat{\alpha}}\nu$ and
$\mu\approx\nu$;

\end{enumerate}
and the symmetric conditions of (1) and (2).
\end{theorem}
\begin{proof} Similar to the corresponding result, Theorem 36, of \cite{FDJY07}. \end{proof}

\begin{lemma}\label{lem:traceequal}
If $\<P,\rho\>\approx \<Q,\sigma\>$, then for any super-operator $\e$ acting on $\h_{\overline{qv(P)}}$, we have $\tr(\e(\rho))=\tr(\e(\sigma))$. In particular, $\tr(\rho)=\tr(\sigma)$. 
\end{lemma}
\begin{proof} Let $S=qv(P)$. From $\<P,\rho\>\approx \<Q,\sigma\>$, we have $\tr_{S} (\rho) = \tr_{S} (\sigma).$
Note that $\e(\tr_{S}(\rho)) = \tr_{S} (\e(\rho))$ since $\e$ acts only on $\h_{\overline{S}}$, and $\tr(\e(\rho))=\tr_{\overline{S}}(\tr_{S} (\e(\rho))).$ The result follows. \end{proof}

As in classical process algebra, the notion of weak bisimulation up to $\approx$ is useful: 

\begin{definition}
A relation $\r\subseteq Con\times Con$ is called a weak bisimulation up to $\approx$ if for any $\<P, \rho\>, \<Q, \sigma\>\in Con$, $\<P, \rho\>\r \<Q, \sigma\>$ implies that  $qv(P)=qv(Q)$, $\tr_{qv(P)} (\rho) = \tr_{qv(Q)} (\sigma)$, and
\begin{enumerate}
\item whenever $\<P,\rho\> \rto{\qc c?q} \mu$, then $\<Q,\sigma\>\Rto{}\rto{\qc c?q} \nu$ for some $\nu$ such that for any trace-preserving super-operator $\e$ acting on $\h_{\overline{qv(\mu) - \{q\}}}$,  $\e(\mu){\r\circ\approx}\e(\nu)$;

\item whenever $\<P,\rho\> \rto{\alpha} \mu$ where $\alpha$ is not a quantum input, then there
exists $\nu$ such that $\<Q,\sigma\>\Rto{\widehat{\alpha}}\nu$ and
$\mu{\r\circ\approx}\nu$;

\end{enumerate}
and the symmetric conditions of (1) and (2).
\end{definition}

\begin{lemma}\label{lem:upto}
If $\r$ is a weak bisimulation up to $\approx$, then $\r\subseteq {\approx}$.
\end{lemma}
\begin{proof} Suppose  $\r$ is a weak bisimulation up to $\approx$. We first prove that that 
$\r\circ\approx$ is a weak bisimulation.
Let $\<P,\rho\>\r\circ\approx \<Q,\sigma\>$; that is, there exists $\<R,\eta\>$ such that $\<P,\rho\>\r \<R,\eta\>$ and $\<R,\eta\>\approx \<Q,\sigma\>$. Then $qv(P)=qv(R)=qv(Q)$, and $\tr_{qv(P)} (\rho) =\tr_{qv(R)} (\eta) =\tr_{qv(Q)} (\sigma)$.

Let $\<P,\rho\>\rto{\qc c?q}\mu$. Then $\<R,\eta\>\Rto{}\rto{\qc c?q}\omega$ such that for any trace-preserving super-operator $\e$ acting on $\h_{\overline{qv(\mu) - \{q\}}}$, 
$\e(\mu){\r\circ\approx} \e(\omega)$. We further derive from Lemma~\ref{lem:exttrans} that $\<Q,\sigma\>\Rto{}\rto{\qc c?q} \nu,$
and for any trace-preserving super-operator $\f$ acting on $\h_{\overline{qv(\omega) - \{q\}}}$, 
$\f(\omega){\approx} \f(\nu)$. Note that $qv(\mu)=qv(\omega)$. We have
$\e'(\mu){\r\circ\approx} \e'(\nu)$ for any trace-preserving super-operator $\e'$ acting on $\h_{\overline{qv(\mu) - \{q\}}}$, by Lemma~\ref{lem:weight}. 

Let $\<P,\rho\>\rto{\alpha}\mu$ for some $\alpha$ not a quantum input. Then $\<R,\eta\>\Rto{\widehat{\alpha}}\nu$ such that $\mu{\r\circ\approx} \nu$. Furthermore, from $\<R,\eta\>\approx \<Q,\sigma\>$  we have $\<Q,\sigma\>\Rto{\widehat{\alpha}}\omega$ such that $\nu{\approx} \omega$, by Lemma~\ref{lem:exttrans}. So we have $\mu{\r\circ\approx} \omega$ from Lemma~\ref{lem:weight}.

The symmetric form when $\<Q,\sigma\>$ performs an action can be similarly proved. So $\r\circ\approx$ is a weak bisimulation; that is, ${\r \circ\approx}\subseteq {\approx}$. Then the result holds by noting that the identity relation is a trivial weak bisimulation.  
\end{proof}

\subsection{Weak bisimilarity congruence}

We now turn to prove the congruence properties of weak bisimilarity. First, we show
that the weak bisimilarity for configurations is preserved by all static constructors.

\begin{theorem}\label{thm:wbcpreserve} If $\<P,\rho\>\approx \<Q,\sigma\>$ then
\begin{enumerate}
\item
$\<P\| R,\rho\>\approx \<Q\| R,\sigma\>$;
\item
$\<P[f],\rho\>\approx \<Q[f],\sigma\>$;
\item
$\<P\backslash L,\rho\>\approx \<Q\backslash L,\sigma\>$;
\item
$\<\iif\ b\ \then\ P,\rho\>\approx \<\iif\ b\ \then\ Q,\sigma\>$.
\end{enumerate}
\end{theorem}\begin{proof} Let us prove (1); other cases are simpler.
Let
\begin{eqnarray*}
\r&=&\{(\<P\|R, \e(\rho)\>,\<Q\|R,\e(\sigma)\>)\ :\
\<P, \rho\>\approx \<Q, \sigma\>, \\
& & \ \ \ \mbox{ and }\e \mbox{ is a trace-preserving super-operator acting on }
\h_{\overline{qv(P)}}\}.
\end{eqnarray*}
It suffices to show that $\r$ is a weak bisimulation. 
Suppose $(\c, \d)\in \r$  where $\c=\<P\|R, \e(\rho)\>$ and $\d= \<Q\|R,\e(\sigma)\>$ for some $\<P,\rho\>\approx \<Q,\sigma\>$, and $\e$ is a trace-preserving super-operator acting on $\h_{\overline{qv(P)}}$. Then $qv(P)=qv(Q)$ and
$\tr_{qv(P)} (\rho) = \tr_{qv(Q)} (\sigma)$ by Theorem \ref{thm:bisimulation}. Thus $qv(P\|R)=qv(Q\|R)$ and
$$\tr_{qv(P\|R)} (\e(\rho)) = \tr_{qv(Q\|R)} (\e(\sigma)).$$
Let $\<P\|R, \e(\rho)\> \rto{\alpha}\mu$ for some $\alpha$ and $\mu$. There are three cases to consider.
{\renewcommand{\theenumi}{\Roman{enumi}}
\renewcommand{\labelenumi}{\theenumi:}
\renewcommand{\theenumii}{\roman{enumii}}
\renewcommand{\labelenumii}{\theenumii:}
\begin{enumerate}
\item The transition is caused by $P$ solely. We need to further consider two subcases:

\begin{enumerate}

\item $\alpha=\qc c?q$ is a quantum input. Then there exists a transition $\<P, \rho\>\rto{\qc c?q} \<P',\rho\>$ and
$\mu=\<P'\|R,\e(\rho)\>$. By the assumption that $\<P, \rho\>\approx \<Q, \sigma\>$, we have
$$\<Q, \sigma\>\Rto{}\boxplus_{i\in I} p_i\bullet \<Q_i',\sigma_i\>\rto{\qc c?q}\boxplus_{i\in I} p_i\bullet \<Q_i,\sigma_i\>$$ such that for any trace-preserving super-operator $\f$ acting on $\h_{\overline{qv(P')-\{q\}}}$, 
\begin{equation}\label{eq:tmp06271}
\<P', \f(\rho)\>\approx \<Q_i, \f(\sigma_i)\>
\end{equation}
holds for any $i\in I$. Then
$\<Q,\e(\sigma)\>\Rto{}\boxplus_{i\in I} p_i\bullet \<Q_i',\e(\sigma_i)\>$
by Lemma \ref{lem:superoperator2}(3), from which we further derive  $$
\<Q,\e(\sigma)\> \Rto{}\rto{\qc c?q} \boxplus_{i\in I} p_i\bullet \<Q_i,\e(\sigma_i)\>$$ and
$$\<Q\|R,\e(\sigma)\>\Rto{}\rto{\qc c?q} \nu=\boxplus_{i\in I} p_i\bullet \<Q_i\|R,\e(\sigma_i)\>.$$ 

For any trace-preserving super-operator $\f'$ acting on $\h_{\overline{qv(P'\|R)-\{q\}}}$,  we obtain from Lemma~\ref{lem:qvchange} that the composite map $\f'\circ\e$ is a trace-preserving super-operator acting on $\h_{\overline{qv(P')-\{q\}}}$. Now using Eq.(\ref{eq:tmp06271}) we have $$\<P', \f'(\e(\rho))\>\approx \<Q_i, \f'(\e(\sigma_i))\>,$$ and thus
$\<P'\|R, \f'(\e(\rho))\>\r\<Q_i\|R, \f'(\e(\sigma_i))\>.$ That is, $\f'(\mu)\r \f'(\nu)$ as required.

\item $\alpha$ is not a quantum input. Then there exists a transition $\<P, \rho\>\rto{\alpha}\mu_1= \boxplus_{i\in I}
p_i\bullet \<P_i,\rho_i\>$ and $\mu=\boxplus_{i\in I} p_i\bullet
\<P_i\|R,\e(\rho_i)\>$ by Lemma \ref{lem:superoperator}(3). From the assumption that $\<P,\rho\>\approx \<Q, \sigma\>$, we have $$\<Q,
\sigma\>\Rto{\widehat{\alpha}}\nu_1=\boxplus_{j\in J} q_j\bullet \<Q_j,\sigma_j\>$$ and $\mu_1\approx\nu_1$ by Theorem \ref{thm:bisimulation}. 
Noting that $\e$ is a trace-preserving super-operator on $\h_{\overline{qv(Q)}}$, we have
$\<Q,\e(\sigma)\>\Rto{\widehat{\alpha}} \boxplus_{j\in J} q_j\bullet \<Q_j,\e(\sigma_j)\>$ by Lemma \ref{lem:superoperator2}(3). 
So it holds that $$\<Q\|R,\e(\sigma)\>\Rto{\widehat{\alpha}}\nu=\boxplus_{j\in J}
q_j\bullet \<Q_j\|R,\e(\sigma_j)\>.$$

Now for each $i\in I$ and $j\in J$, $\<P_i,\rho_i\>\approx\<Q_j,\sigma_j\>$ implies 
$\<P_i\|R,\e(\rho_i)\>\r\<Q_j\|R,\e(\sigma_j)\>$ since
from Lemma \ref{lem:qvchange}, $\e$ is also a trace-preserving super-operator acting on $\h_{\overline{qv(P_i)}}$.
Thus we have $\mu\r\nu$ by Lemma~\ref{lem:weightdec},
by noting that
 $\mu_1\approx \nu_1$.

\end{enumerate}

\item The transition is caused by $R$ solely. We also need to further consider three subcases:
\begin{enumerate}
\item $\alpha=\qc c?q$ is a quantum input where $q\not\in qv(P)$. Then we have $\<R, \e(\rho)\>\rto{\qc c?q} \<R',\e(\rho)\>$ for some $R'$, and
$\mu=\<P\|R',\e(\rho)\>$. Thus $\<R, \e(\sigma)\>\rto{\qc c?q} \<R',\e(\sigma)\>$. By inference rule \textbf{Inp-Int},  we have
$$\<Q\|R,\e(\sigma)\>\rto{\qc c?q} \<Q\|R',\e(\sigma)\>$$
since $q\not\in qv(Q)$.
Now for any trace-preserving super-operator $\f$ acting on $\h_{\overline{qv(P\|R') - \{q\}}}$,  the composite map $\f\circ\e$ is a trace-preserving super-operator acting on $\h_{\overline{qv(P)}}$ from the fact that $qv(P\|R')-\{q\}\supseteq qv(P)-\{q\}=qv(P)$. Thus
$$\<P\|R', \f(\e(\rho))\>\r\<Q\|R', \f(\e(\sigma))\>$$
from the definition of $\r$.

\item $\alpha=\tau$, and the transition is caused by a measurement prefix $M[\widetilde{q};x]$ where $M=\sum_{i\in I}\lambda_{i}|\psi_{i}\>\<\psi_{i}|$. Then we have
$\<R, \e(\rho)\>\rto{\alpha}\sum_{i\in I}
p_{i} \<R_{i}, \f_{i}(\e(\rho))\>$ where $p_{i}=\tr(|\psi_{i}\>_{\widetilde{q}}\<\psi_{i}|\e(\rho))$, $R_{i}=R\{\lambda_{i}/x\}$, 
$\f_{i}$ is the trace-preserving super-operator which sets the $\widetilde{q}$ systems to $ |\psi_{i}\>\<\psi_{i}|$; that is,
$$\f_{i}(\eta) = \sum_{k\in I} |\psi_{i}\>_{\widetilde{q}}\<\psi_{k}| \eta |\psi_{k}\>_{\widetilde{q}}\<\psi_{i}|$$ for any $\eta\in\d(\h)$,
and $\mu = \sum_{i\in I} p_{i} \<P\|R_{i}, \f_{i}(\e(\rho))\>$. We further derive that
$$\<Q\| R, \e(\sigma)\>\rto{\alpha}\nu=\sum_{i\in I}
q_{i} \<Q\| R_{i}, \f_{i}(\e(\sigma))\>$$
with $q_{i}=\tr(|\psi_{i}\>_{\widetilde{q}}\<\psi_{i}|\e(\sigma)) $. 

Notice that for any
$i$, the composite map $\e_{i}\circ \e$ is a super-operator acting on $\h_{\overline{qv(P)}}$ where 
$\e_{i}(\eta) = |\psi_{i}\>_{\widetilde{q}}\<\psi_{i}|\eta |\psi_{i}\>_{\widetilde{q}}\<\psi_{i}|$ for any $\eta\in\d(\h)$. It follows that $p_i=q_i$ from Lemma \ref{lem:traceequal}. Furthermore, we have
$$(\<P\|R_i, \f_{i}(\e(\rho))\>,\<Q\|R_i, \f_{i}(\e(\sigma))\>)\in\r$$
since $\f_{i}\circ\e$ is a trace-preserving super-operator
acting on $\h_{\overline{qv(P)}}$. 
Then it follows that
$\mu\r \nu$ from Lemma~\ref{lem:weightdec}.

\item $\alpha$ is not a quantum input and the transition is not caused by a measurement.
Then there exists a transition $\<R, \e(\rho)\>\rto{\alpha}
\<R', \f(\e(\rho))\>$ where $\f$ is a trace-preserving super-operator on
$\h_{qv(R)}$, and $\mu=\<P\|R', \f(\e(\rho))\>$. We also have $\<R, \e(\sigma)\>\rto{\alpha}\<R', \f(\e(\sigma))\>$.
 Thus
$\<Q\|R, \e(\sigma)\>\rto{\alpha}\<Q\|R', \f(\e(\sigma))\>,$ and
$$(\<P\|R', \f(\e(\rho))\>,\<Q\|R', \f(\e(\sigma))\>)\in\r$$
since $\f\circ\e$ is a trace-preserving super-operator acting on $\h_{\overline{qv(P)}}$.
\end{enumerate}

\item  The transition is caused by a communication
between $P$ and $R$. Without loss of generality, we assume that
$$\<P,\rho\>\rto{\qc c?q} \<P',\rho\>,\ \ \  \<R,\rho\>\rto{\qc c!q}\<R',\rho\>,$$  and $\mu=\<P'\|R',\e(\rho)\>$.
Other cases are simpler. Then $q\not\in qv(P)$ by the validity of $P\|R$, and $\<R, \eta\>\rto{\qc c!q} \<R',\eta\>$ for any
$\eta\in \dh$.
From the assumption that $\<P, \rho\>\approx \<Q, \sigma\>$, we have
$$\<Q, \sigma\>\Rto{}\boxplus_{i\in I} p_i\bullet \<Q_i',\sigma_i\>\rto{\qc c?q}\boxplus_{i\in I} p_i\bullet \<Q_i,\sigma_i\>$$  such that for any $i\in I$ and any trace-preserving super-operator $\f$ acting on $\h_{\overline{qv(P')-\{q\}}}$, it holds that
$\<P', \f(\rho)\>\approx \<Q_i, \f(\sigma_i)\>$. In particular, we have
\begin{equation}\label{eq:tmp06272}
\<P', \e(\rho)\>\approx \<Q_i, \e(\sigma_i)\>
\end{equation} 
since $qv(P)\supseteq qv(P')-\{q\}$.
Noting that $\e$ is a trace-preserving super-operator on $\h_{\overline{qv(Q)}}$, we have
$\<Q,\e(\sigma)\>\Rto{}\boxplus_{i\in I} p_i\bullet \<Q_i',\e(\sigma_i)\>$
 by Lemma \ref{lem:superoperator2}(3), from which we derive further 
$$\<Q,\e(\sigma)\> \Rto{}\rto{\qc c?q} \boxplus_{i\in I} p_i\bullet \<Q_i,\e(\sigma_i)\>,
$$ and
$$\<Q\|R,\e(\sigma)\>\Rto{}\rto{\tau} \nu=\boxplus_{i\in I} p_i\bullet \<Q_i\|R',\e(\sigma_i)\>.$$ 
Furthermore, for any $i\in I$,  we have
$$(\<P'\|R',\e(\rho)\>,\<Q_i\|R',\e(\sigma_i)\>)\in \r$$ by Eq.(\ref{eq:tmp06272}). That is, $\mu\r\nu$ as required.
\end{enumerate}
}

The symmetric form when $\<Q\|R,\e(\sigma)\>\rto{\alpha}\nu$ can be similarly proved. 
So $\r$ is a
weak bisimulation on $Con$. The result follows by noting that the
identity transformation is also a trace-preserving super-operator on
$\h_{\overline{qv(P)}}$. 
\end{proof}

From Theorem \ref{thm:wbcpreserve}, the superdense coding protocol and teleportation protocol presented in Section 3 are still valid in any quantum process context which consists only of parallel composition, relabeling, restriction, and conditional.

Similar to classical value-passing CCS, the weak bisimilarity for quantum processes is preserved by all
the combinators of qCCS except for summation.

\begin{theorem}\label{thm:wbpreserve} If $P\approx Q$ then
\begin{enumerate}
\item
$a.P\approx a.Q$, $a\in \{\tau,c?x,c!e,\qc c?q,\qc
c!q,\e[\widetilde{q}],M[\widetilde{q};x]\}$;
\item
$P\| R\approx Q\|R$;
\item
$P[f]\approx Q[f]$;
\item
$P\backslash L\approx Q\backslash L$;
\item
$\iif\ b\ \then\ P \approx \iif\ b\ \then\ Q$.
\end{enumerate}
\end{theorem}
\begin{proof} The proof for (1) is similar to Theorem 38 of \cite{FDJY07}, and (2)-(5) are direct from Theorem~\ref{thm:wbcpreserve}. 
\end{proof}

\subsection{Congruent equivalence of quantum processes}

As in classical process algebra, the weak bisimilarity is not preserved by the summation combinator. To deal with this problem, we introduce the notion of equality between quantum processes based on $\approx$.

\begin{definition}\label{def:congruence}
Two configurations $\<P, \rho\>$ and $\<Q, \sigma\>$ are said to be equal, denoted by $\<P, \rho\>\simeq \<Q, \sigma\>$,  if $qv(P)=qv(Q)$, $\tr_{qv(P)} (\rho) = \tr_{qv(Q)} (\sigma)$, and
\begin{enumerate}

\item whenever $\<P,\rho\> \rto{\qc c?q} \mu$, then $\<Q,\sigma\>\Rto{}\rto{\qc c?q} \nu$ for some $\nu$ such that for any trace-preserving super-operator $\e$ acting on $\h_{\overline{qv(\mu) - \{q\}}}$,  $\e(\mu)\approx\e(\nu)$;

\item whenever $\<P,\rho\> \rto{\alpha} \mu$ where $\alpha$ is not a quantum input, then there
exists $\nu$ such that $\<Q,\sigma\>\Rto{\alpha}\nu$ and
$\mu\approx\nu$;

\end{enumerate}
and the symmetric conditions of (1) and (2).

\end{definition}

The only difference between the definitions of $\approx$ and $\simeq$ is that in the latter the $\Rto{\widehat{\alpha}}$ 
transition in clause (2) is replaced by $\Rto{\alpha}$; that is, the matching actions for a $\tau$-move have to be at least one $\tau$-move.

Furthermore, we lift the definition of equality to quantum processes as follows. For $P ,Q\in qProc$, $P\simeq Q$ if and only if for any quantum state $\rho\in \d(\h)$ and any indexed
set $\widetilde{v}$ of classical values, $\< P\{\widetilde{v}/\widetilde{x}\}, \rho\>\simeq \<Q\{\widetilde{v}/\widetilde{x}\}, \rho \>$ where $\widetilde{x}=fv(P)\cup fv(Q)$.

It is worth noting that all the weak bisimulation relations proved in the examples of previous sections are also valid when $\approx$ is replaced by $\simeq$. The following properties are easy to show:
 \begin{theorem}\label{thm:tmp030411}
 \begin{enumerate}
\item $\simeq$ is an equivalence relation;
 \item $P\sim Q$ implies $P\simeq Q$, and $P\simeq Q$ implies $P\approx Q$;
 \item If $P\approx Q$ then $a.P\simeq a.Q$ for $a\in \{\tau,c?x,c!e,\qc c?q,\qc
c!q,\e[\widetilde{q}],M[\widetilde{q};x]\}$;
\item $P\simeq Q$ if and only if $P + R \approx Q + R$ for all $R \in qProc$.
\end{enumerate}
 \end{theorem}

Now we prove that the equality relation is preserved by all process constructors of qCCS.

\begin{theorem}\label{lem:eqpreserve} If $P\simeq Q$ then
\begin{enumerate}
\item
$a.P\simeq a.Q$, $a\in \{\tau,c?x,c!e,\qc c?q,\qc
c!q,\e[\widetilde{q}],M[\widetilde{q};x]\}$;
\item
$P+R\simeq Q+R$;
\item
$P\| R\simeq Q\|R$;
\item
$P[f]\simeq Q[f]$;
\item
$P\backslash L\simeq Q\backslash L$;
\item
$\iif\ b\ \then\ P \simeq \iif\ b\ \then\ Q$.
\end{enumerate}
\end{theorem}
\begin{proof} (2) is direct from Theorem~\ref{thm:tmp030411} (4). Others are similar to the proofs of corresponding results for weak bisimilarity. 
\end{proof}

We now turn to examine the properties of the congruent equivalence $\simeq$ under recursive definitions. 
\begin{definition}
Let $E$ and $F$ be process expressions containing at most process
variables $\{X_i(\widetilde{q_i}) : i\in I\}$. Then $E$ and $F$ are equal, denoted by $E\simeq F$, if for all family $\{P_i : i\in I\}$ of quantum processes with $qv(P_i)\subseteq \widetilde{q_i}$,
we have $$E\{P_i/X_i(\widetilde{q_i}) : i\in I\}\simeq F\{P_i/X_i(\widetilde{q_i}) : i\in I\}.$$
\end{definition}

The next theorem shows that $\simeq$ is also preserved by recursive definitions.

\begin{theorem}\label{thm:eqpreserve}
\begin{enumerate}
\item
If $A(\widetilde{q})\define P$, then $A(\widetilde{q})\simeq P$;
\item
Let $\{E_i : i\in I\}$ and  $\{F_i : i\in I\}$ be two families of process expressions containing at most process
variables $\{X_i(\widetilde{q_i}): i\in I\}$, and $E_i\simeq F_i$ for each $i\in I$. If $\{A_i(\widetilde{q_i}) : i\in I\}$ and  $\{B_i(\widetilde{q_i}) : i\in I\}$ be two families of process constants such that
\begin{eqnarray*}
A_i(\widetilde{q_i})&\define&  E_i\{A_j(\widetilde{q_j})/X_j(\widetilde{q_j}) : j\in I\}\\
B_i(\widetilde{q_i})&\define&  F_i\{B_j(\widetilde{q_j})/X_j(\widetilde{q_j}) : j\in I\},
\end{eqnarray*} then $A_i(\widetilde{q_i})\simeq B_i(\widetilde{q_i})$ for all $i\in I$.
\end{enumerate}
\end{theorem}
\begin{proof} (1) is obvious. For (2), we only prove the special case where $|I|=1$ and for any $i\in I$, $\widetilde{q_i}=\emptyset$. That is, we prove
$A\simeq B$ assuming that $qv(A)=qv(B)=\emptyset$,
$A\define  E(A)$ and $B\define  F(B)$ where  $E$ and  $F$ are process expressions containing process
variable  $X$ with $qv(X)=\emptyset$,  and $E\simeq F$.

Let $$\r=\{(\<G(A), \rho\>, \<G(B), \rho\>) : \rho\in \dh, G \mbox{ contains at most process variable } X\}.$$
Obviously, for any $\<G(A), \rho\>\r\<G(B), \rho\>$, we have $qv(G(A))=qv(G(B))$ and $\tr_{qv(G(A))}(\rho)=\tr_{qv(G(B))}(\rho)$. 
Similar to Propositions 4.12 and 7.8 of \cite{Mi89}, we can prove the following properties by induction on the depth of the inference by which the action $\<G(A),\rho\> \rto{\alpha} \mu$ is inferred:

(i) whenever $\<G(A),\rho\> \rto{\qc c?q} \mu$, then $\<G(B),\rho\>\Rto{}\rto{\qc c?q} \nu$ such that for any trace-preserving super-operator $\e$ acting on $\h_{\overline{qv(\mu) - \{q\}}}$, 
$\e(\mu){\r\circ\approx} \e(\nu)$;

(ii) whenever $\<G(A),\rho\> \rto{\alpha} \mu$ where $\alpha$ is not a quantum input, there
exists $\nu$ such that $\<G(B),\rho\>\Rto{\alpha}\nu$ and
$\mu{\r\circ\approx}\nu$.

Only one case deserves elaboration: when $G=G_1\|G_2$ and $\<G(A),\rho\>\rto{\tau}\<P', \rho\>$ is caused by 
$$\<G_1(A),\rho\>\rto{\qc c?q}\<P_1', \rho\>\ \ \mbox{   and   }\ \ \<G_2(A),\rho\>\rto{\qc c!q}\<P_2', \rho\>$$ where $P'=P_1'\| P_2'$. By induction, we have
$$\<G_1(B),\rho\>\Rto{}\rto{\qc c?q}\boxplus_{i\in I}p_i \bullet\<Q_1^i, \f'_i(\rho)\>$$
where $\f'_i$ is a trace-preserving super-operator acting on $qv(G_1)$ (here Lemma~\ref{lem:superoperator2}(2) is used for the $\Rto{}$ transition), and for any trace-preserving super-operator $\e$ on $\h_{\overline{qv(P_1')-\{q\}}}$ and any $i\in I$, it holds 
\begin{equation}\label{eq:tmp071102}
\<P_1',\e(\rho)\>\r\<Q_1',\e(\rho)\>\approx\<Q_1^i, \e(\f'_i(\rho)\>.
\end{equation}
 Thus $P_1'=H_1(A)$ and $Q_1'=H_1(B)$ for some $H_1$ containing only process variable $X$.
 
Also by induction, we have
$$\<G_2(B),\rho\>\Rto{\qc c!q}\boxplus_{j\in J}q_j \bullet\<Q_2^j, \f_j(\rho)\>$$ where $\f_j$ is a trace-preserving super-operator acting on $qv(G_2)$, and for any $j\in J$, 
\begin{equation}\label{eq:tmp071103}
\<P_2', \rho\>\r\<Q_2', \rho\>\approx\<Q_2^j, \f_j(\rho)\>.
\end{equation}
Thus $P_2'=H_2(A)$ and $Q_2'=H_2(B)$ for some $H_2$ containing only process variable $X$.

Now by inference rule \textbf{Q-Com}, and noting that $\f_i'$ and $\f_j$ commute for any $i\in I$ and $j\in J$ since $qv(G_1)\cap qv(G_2)=\emptyset$, we derive that
$$\<G(B),\rho\>\Rto{\tau}\boxplus_{i\in I}\boxplus_{j\in J}p_iq_j \bullet\<Q_1^i\|Q_2^j, \f_j(\f_i'(\rho))\>.$$

Now we calculate that for any $i\in I$ and $j\in J$,
\begin{eqnarray*}
\<P_1'\| P_2',\rho\> &= & \<(H_1\| H_2)(A), \rho\>\\
&\r& \<(H_1\| H_2)(B), \rho\> \hspace{2.8em} \mbox{ By definition}\\
& =& \<Q_1'\| Q_2', \rho\>\\
&\approx&  \<Q_1'\| Q_2^j,\f_j(\rho)\>\hspace{3.3em} \mbox{ By Eq.(\ref{eq:tmp071103}) and Theorem \ref{thm:wbcpreserve}}\\
&\approx &    \<Q_1^i\|Q_2^j, \f_j(\f_i'(\rho))\>\hspace{1.5em} \mbox{ By Eq.(\ref{eq:tmp071102}), Lemma~\ref{lem:qvchange}, and Theorem \ref{thm:wbcpreserve}}.
\end{eqnarray*}

Similarly, we can prove the symmetric forms of (i) and (ii) for $\<G(B),\rho\> \rto{\alpha}\nu$. Then $\r$ is a weak bisimulation up to $\approx$, and so $\r{\subseteq} \approx$ by Lemma \ref{lem:upto}. Now from (i) and (ii) again, we have $\<G(A),\rho\>\simeq\<G(B),\rho\>$. Taking $G=X$ and noting the arbitrariness of $\rho$, we have $A\simeq B$. 
\end{proof}

Finally, the uniqueness of solutions of equations can be proved for process expressions in qCCS, in the sense of $\simeq$.

\begin{definition}
Given a process variable $X(\widetilde{q})$ and a process expression $E$, we say 
\begin{itemize}
\item $X(\widetilde{q})$ is sequential in $E$ if every subexpression of $E$ which contains $X(\widetilde{q})$, excluding $X(\widetilde{q})$ itself, is of the form $a.F$, $\sum_{i\in I} F_i$, or $\iif\ b\ \then\ F$;
\item $X(\widetilde{q})$ is guarded in $E$ if each occurrence of $X(\widetilde{q})$ is within some subexpression $a.F$ of $E$ where $a$ is a (classical or quantum) input or output.
\end{itemize}
\end{definition}

We also say that $E$ is sequential (resp. guarded) if each process variable is sequential (resp. guarded) in $E$.

\begin{lemma}\label{lem:gs}
Let $G$ be guarded and sequential, and contain at most process variables $\widetilde{X}$. If $\<G(\widetilde{P}),\rho\>\rto{\alpha}\boxplus_{i\in I} p_i\bullet \<P_i', \rho_i\>$. Then there exist sequential process expressions $\{H_i : i\in I\}$, containing at most process variables $\widetilde{X}$, such that $P_i'=H_i(\widetilde{P}_\alpha)$ for each $i$, and for any $\widetilde{Q}$, $\<G(\widetilde{Q}),\rho\>\rto{\alpha}\boxplus_{i\in I} p_i\bullet \<H_i(\widetilde{Q}_\alpha), \rho_i\>$. 
Here $$\widetilde{P}_\alpha=\left\{%
\begin{array}{ll}
  \widetilde{P}\{r/q\} \mbox{ for some }q\in qv(\widetilde{P}), & \mbox{ if } \alpha=\qc c?r \\
  \widetilde{P}\{v/x\}  \mbox{ for some }x\in fv(\widetilde{P}), & \mbox{ if $\alpha= c?v$ or $\alpha=\tau$ is caused by a measurement}  \\
  \widetilde{P}, & \mbox{ otherwise} 
\end{array}%
\right.
$$ and $\widetilde{Q}_\alpha$ is defined similarly.
Moreover, if $\alpha=\tau$, then $H_i$ is guarded.
\end{lemma}
\begin{proof} Similar to Lemma 7.12 of \cite{Mi89}. 
\end{proof}

\begin{theorem}\label{thm:uniqueness}
Let $\{E_i : i\in I\}$ be a family of process expressions containing at most process
variables $\{X_i(\widetilde{q_i}): i\in I\}$, and each $X_j(\widetilde{q_j})$ is sequential and guarded in each $E_i$. Let $\{P_i : i\in I\}$ and $\{Q_i : i\in I\}$ be two families of quantum processes such that $qv(P_i)\cup qv(Q_i) \subseteq \widetilde{q_i}$ for each $i$, and 
\begin{eqnarray*}
P_i&\simeq&  E_i\{P_j/X_j(\widetilde{q_j}) : j\in I\}\\
Q_i&\simeq&  E_i\{Q_j/X_j(\widetilde{q_j}) : j\in I\},
\end{eqnarray*} then $P_i\simeq Q_i$ for all $i\in I$.
\end{theorem}
\begin{proof} For simplicity, we only prove the case where $|I|=1$ and all the processes contain no free classical or quantum variables. That is, we prove
$P\simeq Q$ assuming that $qv(P)=qv(Q)=\emptyset$,  $fv(P)=fv(Q)=\emptyset$, 
$P\simeq  E(P)$, and $Q\simeq E(Q)$, where $E$ contains at most process
variable  $X$.

Let 
\begin{eqnarray*}
\r&=&\{(\<M, \rho\>, \<N,\sigma\>) : \<M, \rho\>\approx \<H(P),\eta\> \mbox { and } \<N, \sigma\>\approx \<H(Q),\eta\> \\
&&\hspace{3em} \mbox{ for some } \eta\in\dh,  \mbox{ and }H \mbox{ is sequential and contains at most }X \}.
\end{eqnarray*}
We show $\r$ is a weak bisimulation. The proof is somewhat similar to Proposition~7.13 in \cite{Mi89}. We first claim that for any $\<M, \rho\>\r \<N,\sigma\>$,
\begin{equation}\label{eq:tmp071104}
\mbox{If $\<M, \rho\>\Rto{}\mu$, then $\<N,\sigma\>\Rto{}\nu$ such that $\mu\r\nu$}
\end{equation}
Suppose $\<M, \rho\>\Rto{}\mu$. Then $\<H(P),\eta\>\Rto{}\mu_1$, $\mu \approx \mu_1$, from $\<M, \rho\>\approx \<H(P),\eta\>$. By Theorem~\ref{lem:eqpreserve}, we have $H(E(P))\simeq H(P)$, so $\<H(E(P)),\eta\>\Rto{}\mu_2$ such that $\mu_1 \approx \mu_2$.
Note that $X$ is both sequential and guarded in $H(E(P))$. By repeatedly using Lemma~\ref{lem:gs}, we have
$\mu_2=\boxplus_{i\in K}p_i\bullet \<H'_i(P), \rho_i\>$, and $$\<H(E(Q)),\eta\>\Rto{}\nu_2=\boxplus_{i\in K}p_i\bullet \<H'_i(Q), \rho_i\>$$ 
where $H'_i$ is sequential for any $i\in K$. Since $H(E(Q))\simeq H(Q)$ and $\<N, \sigma\>\approx \<H(Q),\eta\>$, we have
$\<H(Q),\eta\>\Rto{}\nu_1$, $\nu_2 \approx \nu_1$, and $\<N,\sigma\>\Rto{}\nu$, $\nu_1 \approx \nu$. Furthermore, it is obvious that $\mu_2 \r\nu_2$ from Lemma~\ref{lem:weightdec}, and then $\mu \r\nu$  by Lemma~\ref{lem:weight} since $\approx\circ \r\circ \approx \subseteq \r$.

Now let $\<M, \rho\>\rto{\alpha}\mu$ where $\alpha\neq \tau$. There are two cases to consider:
\begin{enumerate}
\item $\alpha=\qc c?q$ is a quantum input. Then $\mu=\<M',\rho\>$ for some $M'$. So $\<H(P),\eta\>\Rto{}\rto{\qc c?q}\mu_1$ such that $\e(\mu) \approx \e(\mu_1)$  for any trace-preserving super-operator $\e$ acting on $\h_{\overline{qv(\mu)-\{q\}}}$.
By Theorem~\ref{lem:exttrans} we further have $\<H(E(P)),\eta\>\Rto{}\rto{\qc c?q}\mu_2$  such that $\f(\mu_1) \approx \f(\mu_2)$ for any trace-preserving  super-operator $\f$ acting on $\h_{\overline{qv(\mu_1)-\{q\}}}$. 
Note that $X$ is both sequential and guarded in $H(E(P))$. By repeatedly using Lemma~\ref{lem:gs}, we have
$\mu_2=\boxplus_{j\in J}q_j\bullet \<H'_j(P), \rho_j'\>$, and $$\<H(E(Q)),\eta\>\Rto{}\rto{\qc c?q}\nu_2=\boxplus_{j\in J}q_j\bullet \<H'_j(Q), \rho_j'\>$$ 
where $H'_j$ is sequential for any $j\in J$. Using Theorem~\ref{lem:exttrans} again we have
$\<H(Q),\eta\>\Rto{}\rto{\qc c?q}\nu_1$ such that $\f'(\nu_2) \approx \f'(\nu_1)$ for any  trace-preserving super-operator $\f'$ acting on $\h_{\overline{qv(\nu_2)-\{q\}}}$,
and $\<N,\sigma\>\Rto{}\rto{\qc c?q}\nu$
such that $\e'(\nu_1) \approx \e'(\nu)$ for any  trace-preserving super-operator $\e'$ acting on $\h_{\overline{qv(\nu_1)-\{q\}}}$. Finally, since $qv(\mu)=qv(\mu_1)=qv(\nu_1)=qv(\nu_2)$,  we have $$\g(\mu) \approx \g(\mu_1) \approx \g(\mu_2) \mbox{  and  } \g(\nu_2) \approx \g(\nu_1) \approx \g(\nu)$$for any trace-preserving  super-operator $\g$ acting on $\h_{\overline{qv(\mu)-\{q\}}}$.
Note that by Lemma~\ref{lem:weightdec}, $\g(\mu_2) \r \g(\nu_2)$. Then $\g(\mu) \r \g(\nu)$  from Lemma~\ref{lem:weight} since $\approx\circ \r\circ \approx \subseteq \r$.

\item $\alpha$ is a quantum output or classical input/output. Then $\<H(P),\eta\>\Rto{\alpha}\mu_1$, $\mu \approx \mu_1$, and  $\<H(E(P)),\eta\>\Rto{\alpha}\mu_2$, $\mu_1 \approx \mu_2$. We further break the actions of $\<H(E(P)),\eta\>$ into
 $$\<H(E(P)),\eta\>\Rto{}\rto{\alpha}\mu_3\Rto{}\mu_2.$$
Note that $X$ is both sequential and guarded in $H(E(P))$. By repeatedly using Lemma~\ref{lem:gs}, we have
$\mu_3=\boxplus_{i\in K}p_i\bullet \<H'_i(P), \rho_i\>$, and $$\<H(E(Q)),\eta\>\Rto{}\rto{\alpha}\nu_3=\boxplus_{i\in K}p_i\bullet \<H'_i(Q), \rho_i\>$$ 
where $H'_i$ is sequential. For any $i\in K$, it is obvious that $\<H'_i(P), \rho_i\>\r \<H'_i(Q), \rho_i\>$. So by Eq.(\ref{eq:tmp071104}) we have $\nu_3\Rto{}\nu_2$ such that $\mu_2 \r\nu_2$. We further derive
$\<H(Q),\eta\>\Rto{\alpha}\nu_1$, $\nu_2 \approx \nu_1$ and $\<N,\sigma\>\Rto{\alpha}\nu$, $\nu_1 \approx \nu$. Finally, we have $\mu \r\nu$ from $\mu_2 \r\nu_2$.
\end{enumerate} 

We have proved that $\r$ is a weak bisimulation. In particular,  for any sequential $H$, $H(P)\approx H(Q)$. Since $E$ is guarded and sequential, every occurrence of $X$ is within some subexpression $a.F$ of $E$ where $F$ is also sequential. Then we have $F(P)\approx F(Q)$, and so $a.F(P)\simeq a.F(Q)$. Thus $E(P)\simeq E(Q)$ by Theorem \ref{lem:eqpreserve}. Now the result $P\simeq Q$ follows from $P\simeq E(P)$ and $Q\simeq E(Q)$.
\end{proof}

To illustrate the power of the theorems proved in this section, let us reconsider Example~\ref{exam:ugate}. We will provide another proof for $\u(U)\circ \u(V) \simeq \u(VU)$ using the Expansion law and the uniqueness of solutions of equations. For simplicity, we only consider the special case where $U$ and $V$ are both 1-qubit unitary operators. Recall the definition of $\u(U)\circ \u(V) $ in Example~\ref{exam:qcircuit}:
\begin{eqnarray*}
\u(U)\circ \u(V) &\define& 
(L_s\|\u(U)[\qc e/\qc c, \qc f/\qc d]\| \u(V)[\qc f/\qc c,\qc g/\qc d]\|R_s)\backslash  L
\end{eqnarray*}
where $L=\{c, \qc e,\qc f,\qc g\}$. Then from Theorem~\ref{thm:eqpreserve} (1), and repeatedly using Theorems~\ref{thm:expansion} and \ref{lem:eqpreserve}, we have 
\begin{eqnarray*}
\u(U)\circ \u(V) &\simeq& \qc c?q.\tau.U[q].\tau.V[q].\tau.\qc d!q.\tau.\u(U)\circ \u(V)
\end{eqnarray*}
where the first $\tau$ action is caused by interaction between $L_s$ and $\u(U)[\qc e/\qc c, \qc f/\qc d]$, the second one between  $\u(U)[\qc e/\qc c, \qc f/\qc d]$ and $\u(V)[\qc f/\qc c, \qc g/\qc d]$, the third one between  $\u(V)[\qc f/\qc c, \qc g/\qc d]$ and $R_s$, while the last one between  $R_s$ and $L_s$.

On the other hand, by Theorem~\ref{thm:eqpreserve} (1) we have 
\begin{eqnarray*}
\u(VU) &\simeq& \qc c?q.VU[q].\qc d!q.\u(VU).
\end{eqnarray*}
Now let $X$ be a quantum process variable with $qv(X)=\emptyset$, and 
$$E=\qc c?q.\tau.U[q].\tau.V[q].\tau.\qc d!q.\tau.X \mbox{,}\hspace{2em} F=\qc c?q.VU[q].\qc d!q.X$$ be two quantum process expressions. Then $E$ and $F$ are both sequential and guarded, and $E\simeq F$. So we have  $\u(U)\circ \u(V) \simeq \u(VU)$ from Theorem~\ref{thm:uniqueness}.

\section{Conclusions and further work}
 
In this paper,  we propose a formal model qCCS, which is a quantum extension of classical value-passing CCS, to model and rigorously analyze the behaviors of quantum distributed computing and quantum communication protocols. We define notions of strong/weak bisimulations for quantum processes in qCCS, and prove that they are preserved by various process constructors, including parallel composition where both classical and quantum communication are present. These are the first congruent equivalences for process algebras proposed so far aiming at modeling quantum communicating systems. We also propose an approximate version of strong bisimulation to characterize the distance between two quantum processes based on strong bisimulation, even when they are not strongly bisimilar. Various examples are fully examined to show the expressiveness of qCCS as well as the proof techniques presented in this paper.

Approximate strong bisimulation has been successfully developed in Section 5. A corresponding notion for weak bisimulation seems, however, very difficult to define. A naive trial is to define a relation $\r$ on $Con$ to be a 
$\lambda$-weak bisimulation if for any $\<P, \rho\>, \<Q, \sigma\>\in Con$, $\<P, \rho\>\r \<Q, \sigma\>$ implies that $qv(P)=qv(Q)$, $d[\tr_{qv(P)} (\rho), \tr_{qv(Q)} (\sigma)]\leq \lambda$, and
\begin{enumerate}

\item whenever $\<P,\rho\> \rto{\qc c?q} \mu$, then $\<Q,\sigma\>\Rto{}\rto{\qc c?q} \nu$ for some $\nu$ such that for any trace-preserving super-operator $\e$ acting on $\h_{\overline{qv(\mu) - \{q\}}}$,  $\e(\mu)\r_{\lambda}\e(\nu)$;

\item whenever $\<P,\rho\> \rto{\alpha} \mu$ where $\alpha$ is not a quantum input, then there
exists $\nu$ such that $\<Q,\sigma\>\Rto{\hat{\alpha}}\nu$ and
$\mu\r_{\lambda}\nu$;
\end{enumerate}
and the symmetric conditions of (1) and (2).
To establish a similar result of Lemma~\ref{lem:abisplus}(1), which is the key for the triangle inequality of the derived bisimulation distance, 
we naturally require that if $\<P, \rho\>\r \<Q, \sigma\>$ for some $\lambda$-weak bisimulation $\r$, then whenever $\<P,\rho\> \Rto{\alpha} \mu$ for some $\alpha\neq\tau$, there
exists $\nu$ such that $\<Q,\sigma\>\Rto{\alpha}\nu$ and $\mu\r_{\lambda}\nu$. However, this property does not hold in general. To see this, suppose $\<P, \rho\>\rto{\tau}\mu'\rto{\alpha} \mu$, and $\<Q,\sigma\>\Rto{}\nu'$ is the weak transition  $\<Q,\sigma\>$ takes to match the action $\<P,\rho\> \rto{\tau} \mu'$, and $\mu'\r_{\lambda}\nu'$. 
However,  the conditions that $\mu'\r_{\lambda}\nu'$ and $\mu'\rto{\alpha}\mu$ do not necessarily imply that $\nu'\Rto{\alpha}\nu$ for some $\nu$; they only guarantee that a portion of $\nu$ with the probability weight not less than $1-\lambda$ can perform a weak $\alpha$-action. Furthermore, even such a $\nu$ exists, we can only infer $\mu\r_{2\lambda}\nu$ from $\mu'\r_{\lambda}\nu'$ but not $\mu\r_{\lambda}\nu$ as expected. That is, the imperfection, or error, which is allowed by approximate bisimulation will accumulate during the execution of weak transitions.

Another interesting direction worth researching is to expand the application scope of qCCS to model and analyze the $security$ $properties$ of quantum cryptographic systems. By introducing cryptographic primitives, such as constructors for encryption and decryption, into pi-calculus, the Spi calculus \cite{AG97} has been very successful in cryptographic protocol analysis. We believe that a similar extension of our qCCS will provide tools for analyzing quantum cryptographic protocols such as BB84 quantum key distribution protocol. 

\begin{acks}
The authors thank Prof. Prakash Panangaden and Dr. Yuxin Deng for their generous comments and suggestions. In particular, Yuxin pointed out a flaw in the definition of weak transitions in our POPL paper, and suggested us to adopt the combined version as in Definition~\ref{def:wtran} of the current paper. The reviewers of POPL'11 are also acknowledged for their comments which have improved the presentation and the quality of the paper. 
\end{acks}

\bibliographystyle{acmsmall}

\end{document}